\documentclass[showpacs,prd,floats,aps,twocolumn,eqsecnum,tightenlines,superscriptaddress]{revtex4}
\usepackage{graphicx}

\begin{document}

\preprint{UTBRG-2002-001, AEI-2001-xxx, gr-qc/0102xxx}

\title{Modeling gravitational radiation from coalescing binary black holes}

\author{J. Baker}
\affiliation{Laboratory for High Energy Astrophysics, 
NASA Goddard Space Flight Center, Greenbelt, Maryland 20771}
\author{M. Campanelli}
\affiliation{Department of Physics and Astronomy, 
The University of Texas at Brownsville, Brownsville, Texas 78520}
\affiliation{Albert-Einstein-Institut, Max-Planck-Institut f{\"u}r Gravitationsphysik, Am M\"uhlenberg 1, D-14476 Golm, Germany}
\author{C. O. Lousto}
\affiliation{Department of Physics and Astronomy, 
The University of Texas at Brownsville,
Brownsville, Texas 78520}
\affiliation{Albert-Einstein-Institut, Max-Planck-Institut f{\"u}r Gravitationsphysik, Am M\"uhlenberg 1, D-14476 Golm, Germany}
\affiliation{Instituto de Astronom\'{\i}a y F\'{\i}sica del Espacio--CONICET,
Buenos Aires, Argentina}
\author{R. Takahashi}
\affiliation{Theoretical Astrophysics Center, Dk-2100 K\o benhavn \O , Denmark}


\date{\today}

\begin{abstract}
With the goal of bringing theory, particularly numerical relativity, 
to bear on an astrophysical problem of critical interest to gravitational 
wave observers we introduce a model for coalescence radiation from binary 
black hole systems. 
We build our model using the {\sl Lazarus approach}, a technique that 
bridges far and close limit approaches with full numerical relativity
to solve Einstein equations applied in the truly nonlinear dynamical regime.
We specifically study the post-orbital radiation from a system of equal-mass 
non-spinning black holes, deriving waveforms which indicate strongly 
circularly polarized radiation of roughly 3\% of the system's total energy 
and 12\% of its total angular momentum in just a few cycles.
Supporting this result we first establish the reliability 
of the late-time part of our model, including 
the numerical relativity and close-limit components, 
with a thorough study of waveforms from 
a sequence of black hole configurations varying from previously treated 
head-on collisions to representative target for ``ISCO'' data corresponding 
to the end of the inspiral period.
We then complete our model with a simple treatment for 
the early part of the spacetime
based on a standard family of initial data for binary black 
holes in circular orbit.
A detailed analysis shows strong robustness in the results
as the initial separation of the black holes is increased from $5.0$ to 
$7.8M$ supporting our waveforms  as a suitable 
basic description of the astrophysical radiation from this system.
Finally, a simple fitting of the plunge waveforms is introduced as a
first attempt to facilitate the task of analyzing data from gravitational
wave detectors.
\end{abstract}

\maketitle

\section{Introduction}\label{Sec:Intro}

Black holes may be the most interesting and extraordinary objects
predicted by Einstein's general relativity. Over the last forty years,
a flowering of observational techniques and theoretical consideration
has produced a change in perspective, bringing black holes from
suspected mathematical fictions to the presumptive powerhouses behind
the most energetic astronomical phenomena under study today.  From
quasars and Gamma Ray Bursts (GRB's) to ordinary galactic nuclei and
the final state of sufficiently large stars, black holes seem now
quite common astrophysical objects.
 The characteristic of black holes which has brought
them to such prominence in astrophysics is that when they are not isolated
they are emphatically {\em not black} but produce the
most luminous astrophysical phenomena known.  
One can estimate that a coalescing binary system of two black holes 
releases energy with a peak luminosity of about $10^{-3}c^5/G$, 
$10^{23}$ times the power output of the
Sun, during the very brief period of its merger.

These events have not yet been observed because, lacking matter, the
binary black hole (BBH) system releases its energy purely in the form
of gravitational waves.  To the developing field of gravitational wave
astronomy these are expected to be generally the brightest sources in
the sky. For ground-based gravitational observatories, such as
LIGO~\cite{Abramovici92} and GEO600, followed shortly by
VIRGO~\cite{Bradaschia90} and several resonant bar detectors, all
sensitive to "higher frequencies" in the $10^1$ - $10^3$ Hz band, the
relevant coalescence events have masses on the order of large stars
$10^1$ - $10^3$ $M_\odot$. Such systems may be generated through the
evolution of stellar binaries, or as recent simulations suggest
through many-body interactions in globular
clusters\cite{Zwart99,Zwart99b}. Space based detectors, such as the
proposed NASA-ESA LISA mission~\cite{Hough94b,lisa3} will be sensitive
to a lower frequency band roughly $10^{-4}$ - $10^{0}$ Hz and the
relevant systems are super-massive black hole binaries involving black
holes in galactic cores.

For both classes of observation, the importance of theoretical models
for the expected BBH coalescence is
extraordinary~\cite{Flanagan97b,Damour00a}.  For the current
generation of ground-based interferometers beginning to take
scientific data in the next few months, the anticipated
signal-to-noise ratios are sufficiently small that theoretical
information from BBH coalescence models may be essential to
distinguishing genuine detections from noise events.  For LISA, on the
other hand, the signal-to-noise ratios for super-massive BBH events
should be quite high, up to $10^4$~\cite{Hughes:2001ch}.  
In this case the BBH events may
be so strong and numerous that the ability to make observations of
other weaker sources depends on filtering out the BBH signals
according to accurate model waveforms.  Because the new ground-based
interferometers are already beginning operations and because model
information may be critical even in the developmental stages of the
LISA mission, there is a pressing need for producing a fair model for
the BBH coalescence immediately.  As general relativistic modeling
techniques continue to evolve and improve, there is naturally a
trade-off between the level of confidence we can have in a model and
how long we wait to produce it, but the clear and present need for
some kind of theoretical result, provides a strong argument against
waiting.  Our goal here is to introduce a practical first model which
may be progressively improved as new techniques are developed.
Producing coalescence waveforms from the first usable models opens up
a vital channel of communication between observational and theoretical
efforts that is now crucially required. Of course we expect this
theoretical information to support the observational effort, providing
observers with specific examples of what sort of signal they can
expect. Importantly, though less widely recognized in the
theoretical community, a provisional model allows observational
practicalities to inform the theoretical effort.  Not all theoretical
concerns are equally consequential in practice, and enhanced
interaction with observers through model waveforms will help to
identify which details of the model need the most work.

Clearly the foundation for a theoretical model of the BBH system will
be Einstein's general theory of relativity.  Intensive efforts to
develop numerical codes able to solve the Einstein field equations
using state-of-the-art supercomputers have been underway for more than
a decade now~\cite{Alliance98}.  These efforts have resulted in
successful studies of several phenomena but, the treatment of binary
black hole systems has proven very difficult.
Thus far, the full
numerical treatment of these systems in three dimensions (3D), based
on the traditional $3+1$ decomposition of space and time, has been
severely limited on one hand, because of huge computer memory and
processing requirements, and, on the other, by instabilities related
to the formulation of the evolution equations.  These difficulties
combine to cause codes to fail to be accurate before any useful
gravitational wave information can be directly extracted even for sample
problems such as, distorted black holes, and the so-called
non-axisymmetric `grazing collisions', where the black holes must
start out yet too close to allow a clear astrophysical interpretation
~\cite{Brandt00,Alcubierre00b} (and even for single
distorted black holes~\cite{Baker99a}).

In an earlier letter ~\cite{Baker01b} we presented the first
theoretical estimates for the gravitational radiation waveforms and
energy to be expected from the plunge of orbiting non-spinning binary
black holes starting from an estimate of the innermost 
stable circular orbit (ISCO) as provided by
Cook/Baumgarte~\cite{Cook94,Baumgarte00a}, using effective potential
method applied to Bowen-York initial data.  Such a binary system has also
been considered by Buonanno and Damour ~\cite{Buonanno00a}, who
produce a radiation waveform estimate without solving the full system
of Einstein's equations in the (expected) nonlinear regime, but using
extrapolations of the 2.5 Post-Newtonian order.  In this paper we
apply the {\it Lazarus} approach~\cite{Baker01a}, to develop a functional
astrophysical model for binary black hole coalescence. 

A complete implementation of our approach requires at least five
components.  There are three basic treatment regimes: (i) a Far-Limit
(FL) treatment covering the earliest region of spacetime when the black holes
are still relatively far apart, (ii) a Fully-Nonlinear (FN) approach
to the essentially nonlinear plunge/merger region of the spacetime and
(iii) a Close-Limit (CL) approach to the dynamics of the newly formed
final black hole and the propagation of radiation. In addition to
these are two interfaces, FL-FN and FN-CL, which require a specific
means of propagating the relevant physical information from the end of
one regime to the beginning of the next. Section
\ref{Sec:Laz_overview} provides an overview of our model and a brief
review of the FN, FN-CL, CL techniques which are described in detail
and tested in principle in Ref. \cite{Baker01a}.  In Section
\ref{Sec:P-seq} we apply theses techniques to perform a systematic
study of the waveforms generated by a set of sample problems,  
simulations of equal mass
binary black holes with no intrinsic spin starting from near the ISCO.
This
provides a proving ground for testing our machinery for taking initial
data at the beginning of the FN regime and producing a final waveform
result. The study connects to previously well-studied problems and
builds up to the practical demands of our astrophysical model.
 with
Spinning and unequal mass black holes 
can also be treated within our method and will be published 
in a different paper.

To complete the model we need to provide FL and FL-FN treatments.  As
our approach to this regime has not been treated extensively in
Ref.\ \cite{Baker01a} we provide here a more detailed exposition.  In
Sec.\ \ref{Sec:latestagemodel} we describe our treatment and discuss how this
approach compares with other candidate treatments which might also
seem appropriate concluding that based on kinematics ours is the
best approach to providing modeled astrophysical data to feed into
into the FN part of the calculation available which is so far
sufficiently developed for practical applications.  But a specific
advantage of having a complete model is that we can move beyond these
(often dubious) kinematical arguments and explicitly study the
robustness of the full dynamical model.  We begin this analysis in
Sec. \ref{Sec:QCresults}, with successful robustness test of the FL
and FL-FN part of the model. We also extend
our results to meet Post-Newtonian (PN) calculations in
Sec.~\ref{Sec:PNcomparisons}. We evolve black holes from the 
Post-Newtonian determined parameters for the ISCO.  
Section
\ref{Sec:Analysis} presents summarizes the astrophysically relevant 
 results of our treatment for the coalescence waveforms, and provides
a simple analytical
approximation of our results as a practical representation for the
benefit gravitational wave observers and data analysts.

\section{The model}\label{Sec:Laz_overview}

Under the Lazarus approach our model is specified by providing a
specific implementation for five components (FL, FL-FN, FN,
FN-CL, CL).  A sketch of this combined approach can be outlined by the
following steps: (1) First provide description of the early dynamics
of the system with a FL approach, such as the PN method, which is
appropriate for slowly moving, well-separated black holes, or some
alternative quasi-stationary (QS) method.  (2) Extract critical
information about the late-time configuration of this system in terms
of Cauchy data, $g_{ij}$ and $K_{ij}$, and translate this information
to a corresponding solution of the gravitational initial-value
problem.  (3) Apply a full 3D numerical simulation of Einstein's
equations to generate a numerical spacetime covering the non-linear
interaction region.  The evolution should proceed for long enough so
that the subsequent evolution of the region exterior to the final
single remnant black hole can be well approximated by perturbative
dynamics.  (4) At this point we choose a ``late-time" slice from the
numerically generated spacetime, extract $\psi_4=
C_{\alpha\beta\gamma\delta}n^\alpha\bar{m}^\beta
n^\gamma\bar{m}^\delta$ and $\partial_t\psi_4$, to quantify the
deviation of the numerical spacetime from a Kerr geometry.  Then (5)
evolve via the Teukolsky equation, which governs the dynamics of Kerr
perturbations in the time-domain
\cite{Teukolsky73}, long enough to drive all significant radiation into
the radiation zone where it can be interpreted.

For convenience in discussion we will refer to steps (3-5) through
which radiation waveforms are derived from numerical simulation
initial data as the ``late-stage'' part of the model.  Likewise the
``early-stage'' part of the model encompasses steps (1-2) in which
numerical data are produced from some model of the early part of the
spacetime.  A suitable implementation of the late-stage part of the
model has been presented in Ref.~\cite{Baker01a}.  During its
development, the late-stage implementation has passed many tests,
including the robustness of $\psi_4$ and $\partial_t\psi_4$ against
nonlinear contamination of the Cauchy data~\cite{Lousto99a}.  In a
first letter \cite{Baker00b} our method proved to be
capable of determining radiation waveforms for a model problem, the
head-on collision of two black holes, with accuracy comparable to the
best published 2D numerical results, allowing at the same time a more
direct physical understanding of the collisions and indicating clearly
when non-linear dynamics are important as the final black hole is
formed.  In \cite{Baker01a} we report on internal consistency checks
after including net angular momentum in the system,
e.g. quadratic convergence to vanishing
gravitational radiation, which our method passes, for the evolution of
Kerr initial data. This is a non-trivial test of our procedure since
the computed spurious radiation energy and waveforms will give us a
direct measure of the `internal' error with which we can determine
such quantities. Typically, the levels of spurious radiation we have
found for rotation parameter $a/M=0.8$ is around $10^{-5}M$.

Below, in Sec.~\ref{Sec:P-seq} we supplement these tests with a 
developmental study of the performance of our late-stage implementation 
on a series of test problems which build from the
previous test problems to a fiducial case of practical interest, 
evolution from ``ISCO'' data, as described below.

In this work we will add to the model a simple treatment for the the
early-stage part.  We essentially model the early part of the
spacetime as a stack of spacelike slices, each slice chosen from a
family of solutions of the Einstein initial value problem representing
non-spinning, orbiting black holes.  The solutions on the slices are
further selected by kinematical (effective potential) arguments to
correspond to circular orbits.  This early-stage part of our model is
clearly less developed than what we have implemented for the
late-stage, and it is fair to note at the start that there are many
reasons to pursue improvements for this part of the model.  The key
advantage of our present approach is that it can be applied even
without further development to provide a complete provisional model
which produces coalescence waveforms.  It is precisely in the context
of such a completed model that we have a concrete basis for evaluating
any treatment of steps (1-2).  In Section \ref{Sec:QCseq} we provide a
detailed description of our treatment and introduce some relevant
concerns related to this part of the model.  In particular we consider
the plausibility of our model in comparison with PN calculations.

One of the most attractive characteristics of the Lazarus approach is
that it provides a built-in facility for cross-checking the three
primary treatments underlying the model.  By shifting the location of
the interface regions in the model spacetime, we can effectively
exchange the treatment chosen for that region of spacetime.  This
creates an inherent facility for validating the three evolutionary
model components.  We can cross-check the Close Limit perturbation theory
treatment with Numerical Relativity results for the same region of
spacetime, systematically incorporating a common technique 
for verifying both Close Limit and numerical 
results~\cite{Baker96a,Anninos95g,Baker99a}.  
In \cite{Baker01a} we
described the basis for such comparisons via the FN-CL interface
looking for phase locking and energy plateaux in the waveforms.  With
the addition of the early-stage treatment we can now begin to
practically compare the FL and FN components of the model. Section
\ref{Sec:QCresults} is dedicated to evaluating the FL model by
comparison with numerical simulation.  Although, the model for the FL
is clearly more suspect, comparison of the results are in much better
agreement than expected, validating our simple treatment as a fair
provisional model.

\section{Practical testing of the late-stage model}\label{Sec:latestagemodel}

Our idea for the late-stage part of the model is to use the Lazarus
tools for deriving 
radiation waveforms, energy and 
momentum radiated, 
plunge durations, 
etc, 
resulting from an evolution of binary black holes that have started
close `enough' to the location of the ISCO.  We approach this case
developmentally, beginning with an axisymmetric system of 
ISCO--separation black holes released from rest.  This system has been
studied extensively using 2D numerical simulations and with a previous
simplified version of our model's late-stage
treatment~\cite{Baker00b}.  From this head-on collision case we
gradually add orbital angular momentum to the system to approach the
target ISCO problem.  We call this sequence of initial data sets the
'P-sequence' through which we vary the transverse momentum of the
individual black hole from zero, in the head-on case to our target
$P = P_{ISCO}$.  The waveforms we present here are the culmination of
a number of numerical simulations which were needed to establish both
the effectiveness of our methodology and the plausibility of these
results as a fair astrophysical model.  The first two subsections lay
the groundwork with a description of the particular family of initial
data we are studying, and an overview of our methodology.

\subsection{Initial Data}\label{Sec:ISCOID}

The Cauchy data required to begin a 3+1 simulation of vacuum general
relativity can be given in the form of a metric tensor $g_{ij}$ on an
initial spacelike and an extrinsic curvature tensor $K_{ij}$ which
contains information about how the initial hypersurface is embedded in
the spacetime.
  The gravitational field equations include
four constraints, limiting the choices for $g_{ij}$ and $K_{ij}$.  To
present data for a binary black hole system generally requires (1)
some astrophysical ansatz for plausible field configurations around
(and constituting) the two black holes (2) solving the constraints.

Many conceivable approaches have been proposed
and to varying degrees developed, for the case of black hole initial data.
In this study we apply a variant of the best known approach, known as 
Bowen-York data~\cite{Bowen80}. This approach provides an ansatz for 
describing black hole systems through a set of restrictions on 
the metric $g_{ij}$ and extrinsic $K_{ij}$.
The assumptions are: conformal flatness $g_{ij}=\Phi^4\delta_{ij}$, 
a maximal slice $K_i^i=0$, and a purely longitudinal
form for the extrinsic curvature $K^{ij}=
\Phi^{-10}(\nabla^i V^j+\nabla^j V^i-\frac{2}{3}g^{ij}\nabla_kV^k)$.
Instead of the traditional ``inversion-symmetric'' boundary 
conditions at each of
the black holes, we apply a simplified variant (`puncture' data
~\cite{Brandt97b}) which allows black hole data 
to be represented numerically without
excising the interior regions of the black holes. An advantage of this
class of data is that is has been studied extensively.

A significant feature, expected to exist in the binary interaction,
is the ISCO indicating the closest
possible configuration of black holes at which the {\it conservative} part
of the dynamics allow stable orbits.
Although the field theory nature of the `Full Numerical' approach does
not lend itself naturally to a particle like description of the
dynamics implicit in describing the ISCO, it is possible to construct
particle-like parameterizations of appropriately chosen families of
solutions to the initial value problem.  In these terms, the
kinematical relationships, among the solutions to the initial value
problem provide a description of the system similar to the
conservative particle dynamics, including an estimate for the ISCO.

Different approaches have been developed to identify quasi-circular
orbital configurations and to determine the location and frequency of
the ISCO.  We will use here the results based on the effective
potential method of ~\cite{Cook94} as derived in \cite{Baumgarte00a}
for the puncture construction of black hole initial data. This
provides a space of solutions parameterized by the momenta, spins and
positions of each black hole.  From these solutions one can apply the
Arnowitt-Deser-Misner (ADM) formalism to determine the total energy
$M_{ADM}$ for the slice of spacetime.  If a bare mass is inferred for
each individual black hole from its apparent horizon area, then the
difference $E_{Eff}$ can be used to identify points in the solution
space corresponding to quasi-circular orbits, and ISCO.

\begin{figure}
\begin{center}
\includegraphics[width=2.3in]{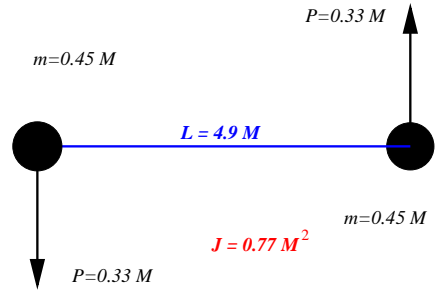}
\end{center}
\caption{Effective potential - ISCO data}
\label{fig:ID}
\end{figure}

For puncture data the ISCO is characterized by the parameters
\begin{eqnarray}
&&L/M=4.9,\quad P/M=0.335,\nonumber \\
&&J/M^2=0.77,\quad m=0.45M.
\end{eqnarray}
where $m$ is the mass of each of the single black holes, $M$ is the
total ADM mass of the binary system,
$L$ is the proper distance between the apparent
horizons, $P$ is the magnitude of the linear momenta (equal but
opposite and perpendicular to the line connecting the holes) and $J$
is the total angular momentum. This configuration is represented
in Fig.\ \ref{fig:ID}

Families of solutions to the initial value problem have been produced
from other sets of assumptions on the fields as well. An approach
inspired from the Kerr-Schild form of a Kerr black hole
\cite{Marronetti00}, provides a more natural model at least when the black 
holes are far apart, but has not yet been studied as a kinematical model for
quasi-circular orbits and ISCO. Another treatment along the lines of the 
maximal/conformally flat formalism 
with no assumption about the longitudinality of 
the extrinsic curvature but with additional restrictions guaranteeing that 
these conditions are initially consistent with the evolution equations, 
is very promising but has not yet been applied to irrotational black hole
systems as considered here~\cite{Grandclement:2001ed,Cook:2001wi,Dain:2002ee}.
Finally, an interesting first step in providing astrophysically realistic 
initial based PN motivated choices for the freely-specifiable portions 
of the data~\cite{Bruegmann02} will be subject of further investigation 
within our approach.

\subsection{methods}\label{Sec:methods}

A detailed discussion of our techniques has been presented in a
previous paper~\cite{Baker01a} so we only briefly review the
techniques here.  We first perform a numerical simulations to cover a
relatively brief period of fundamentally non-linear two-body
interactions to the point where the dynamics of the final black hole
dominate and can be treated linearly.  We then integrate the standard
ADM form of Einstein's equations using the thrice iterated
Crank-Nicholson approach with maximal slicing and vanishing shift,
making extensive use of the publicly available Cactus computational
toolkit~\cite{cactusweb} and Br\"ugmann's multigrid elliptic
solver BAM for initial data and maximal slicing choices.  In designing
these simulations we have focused on the particular needs of our
problem defined within the Lazarus approach, rather than on alternative
techniques for generic applications.  Some of the priorities in this
case are: optimal accuracy for brief simulations in the entire region
outside the final horizon, avoidance of boundary noise, and consistency
with gauge assumptions applied in interfacing to the ``close-limit''
phase of the treatment.  The accuracy requirement tends to favor the
standard ADM over other approaches.  Although the ADM system tends to
be more susceptible to instabilities, it also tends to be more
accurate before these instabilities suddenly arise.

In the cases considered here, these runs would suffer catastrophic
instabilities after $15-20M$ evolution, but the more critical question
for our problem is that the runs be sufficiently accurate in the
convergent regime prior to the instability. In some cases, other
systems~\cite{Alcubierre99e}, seem to allow for significant
improvements in the treatment of boundaries. But at present even these
improved boundary treatments remain problematic, allowing physical
wave content into the computational domain.  For brief, accurate
simulations it is important, and practical to move the boundary far
away, causally separating it from the relevant wave-generation region.
To realize this without undue computational expense we cast the
initial data in specialized {\em fisheye} coordinates which bring a
distant outer boundary to a closer coordinate location without
sacrificing physical resolution in the strong field
regions~\cite{Baker01a}.  Our gauge assumptions favor maximal slicing
in the numerical simulations, and for simplicity without an obvious
alternative, we use a vanishing shift.  Other sophisticated techniques
currently under development in the field, such as black hole excision,
are not so relevant to our narrowly defined numerical simulation
needs.  We effectively excise the interior of the final event horizon
when we make the transition to (purely exterior) close limit
perturbation theory.

In analyzing our results we make extensive use of several special
techniques.  A key tool for interpreting our numerically simulated
spacetimes is the complex-valued `${\cal
S}$-invariant'\cite{Baker00a}.
\begin{equation}
{\cal S}=27{\cal J}^2/{\cal I}^3,
\end{equation}
where ${\cal I}$ and ${\cal J}$ are the two complex curvature
invariants ${\cal I}$ and ${\cal J}$, which 
are essentially the square and cube of the self-dual part,
$\tilde C_{\alpha\beta\gamma\delta}=
C_{\alpha\beta\gamma\delta}+(i/2)\epsilon_{abmn}C^{mn}_{\,\,\,\,cd}$, of
the Weyl tensor:
\begin{equation}
{\cal I}=\tilde C_{\alpha\beta\gamma\delta}
\tilde C^{\alpha\beta\gamma\delta} ~~{\rm and}~~
{\cal J}=\tilde C_{\alpha\beta\gamma\delta}
\tilde C^{\gamma\delta}_{\,\,\,\,\mu\nu}\tilde C^{\mu\nu\alpha\beta}.
\end{equation}
In vacuum general relativity this is the unique (up to rescaling)
spacetime scalar quantity which can be constructed from the components
of the curvature tensor which is not dominated by its fall-off
behavior in an asymptotically flat spacetime.  It is thus a natural
tool for invariant interpretation of spacetime dynamics.  A key
characteristic of the ${\cal S}$-invariant which we take advantage of
here is that is has the exact value, ${\cal S}=1$ in the Kerr geometry
and tends to provide a useful measure of deviations from the Kerr
background.  As a rule of thumb, we adopt the criterion that black
hole perturbation theory is likely to be effective if the real part of
${\cal S}$ differs from unity by less than a factor of two 
(when reaching the maximum) outside of
our estimated location for the horizon.  In these binary black hole
spacetimes, we find the maximal values of ${\cal S}$ on the orbital
($z$) axis.

Our most important means of evaluating the calculations is the
inherent cross-checking between the two evolutionary spacetime models
(close limit and numerical simulation).  By changing the location of
the transition between these approaches in our model spacetime, which
is achieved by varying the evolution time $T$ of the numerical
simulations, we can effectively exchange one treatment for the other.
If the resulting radiation is approximately unchanged, then the two
independent treatments are agreeing with each other and we can be confident
in the results.  In practice we compare the results by separately
examining the magnitude (total radiation energy $E$) and phasing of
the waveforms.  Thus we look at the energy as a function of 
the transition time $T$,
and expect to find a plateau in $E$ once we have
evolved through the most significant regions of non-linear dynamics and
before the inaccuracies in the numerical simulation have grown
dramatically.

\subsection{$P$-Sequence results}\label{Sec:P-seq} 

\begin{figure}
\begin{center}
\includegraphics[width=2.7in]{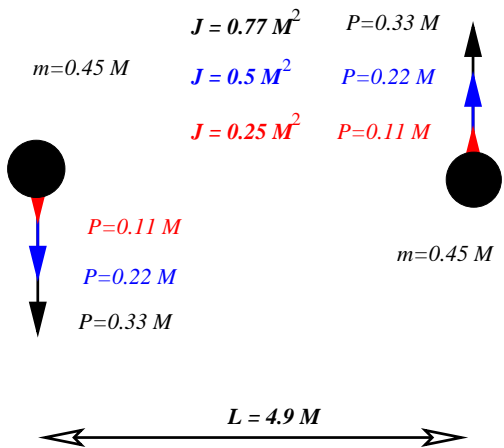}
\end{center}
\caption{Sequence P}
\label{fig:SecP}
\end{figure}

In order to better understand the physics of the plunge we have
designed a set of sequences approaching the ISCO by changing one of
its physical parameters. Many different sequences are possible.  The
`P-sequence' for which we keep the separation constant at
$L=L_{ISCO}=4.9M$, but vary the linear momentum,
$P/P_{ISCO}=0,1/3,2/3,5/6,1$
is particularly interesting because
it connects the well-studied case of head-on collisions from rest to
our target, ISCO data.
Fig.\ \ref{fig:SecP} illustrates the initial configurations studied with
increasing linear momentum from the resting holes to the
ISCO values.

\subsubsection{$P=0$ (Brill-Lindquist initial data)}

The head-on collision of black holes from rest 
has been studied using our techniques
in Ref.~\cite{Baker00b} and compared with $2D$
simulations. Here we revisit this configuration, now not with the
holes lying along the $z-$axis, but along the $y-$axis, and having
switched from Misner to Brill-Lindquist initial data, to serve as a
reference point for the ISCO initial data.  The change in orientation
implies that the radiation is now predominated by is $m=\pm2$ component,
each contributing a factor $3/8$ to the total energy; while
$m=0$ waves have similar time dependence but smaller, contributing
to $1/4$ of the total radiated energy. Note that the reflection
symmetry of the systems considered here implies that the $m=+2$ and
$m=-2$ modes are directly related and contribute the same to the total
energy radiated.  Without loss of information we present only the
$m=+2$ results.

\begin{figure}
\begin{center}
\begin{tabular}{@{}lr@{}}
\includegraphics[width=1.75in]{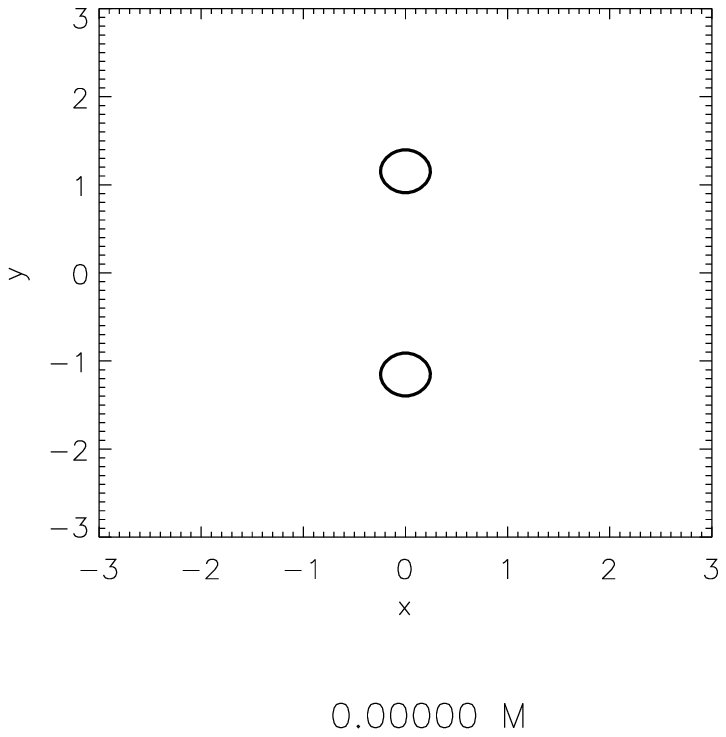} &
\includegraphics[width=1.75in]{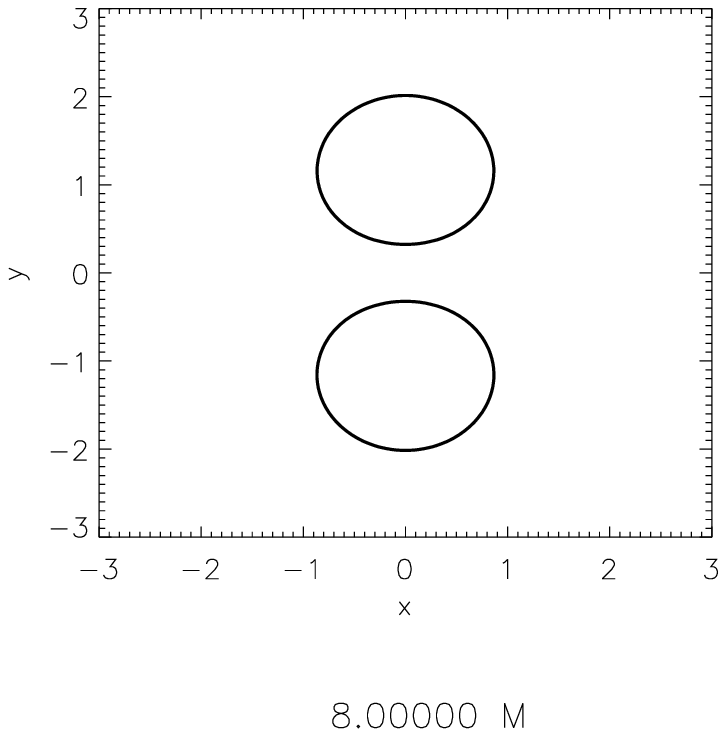}\\
\includegraphics[width=1.75in]{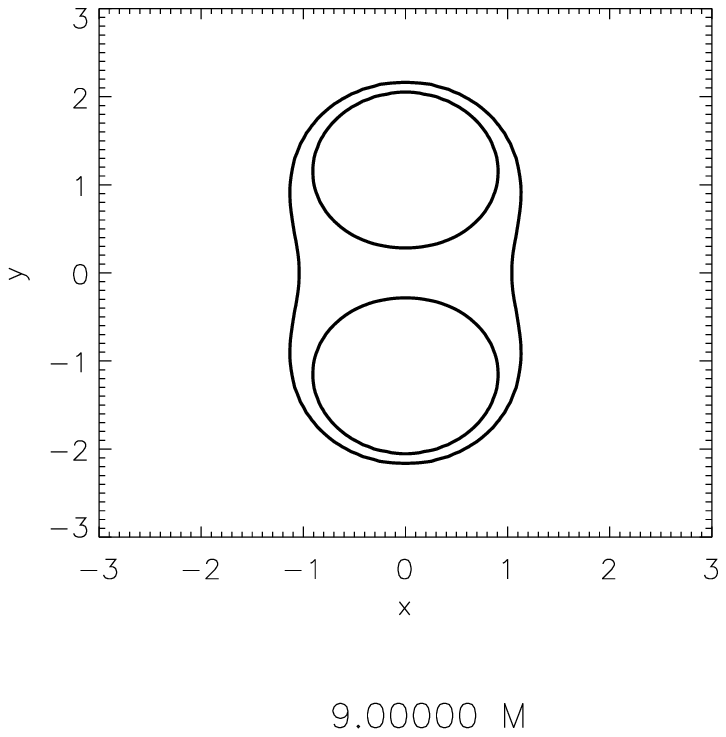} &
\includegraphics[width=1.75in]{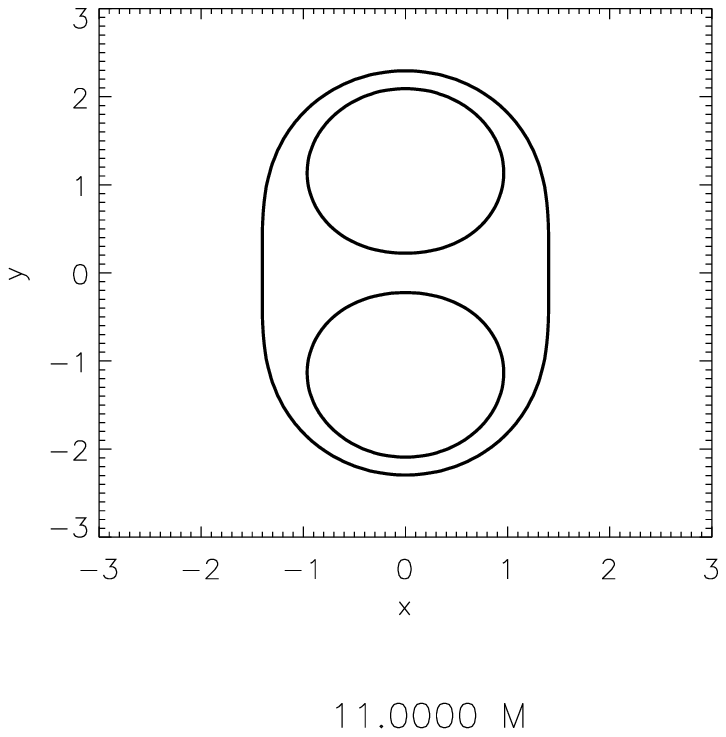}
\end{tabular}
\end{center}
\caption{Apparent horizons for $P_{ID}=0$}
\label{fig:AHSeqP0}
\end{figure}

Fig.\ \ref{fig:AHSeqP0} shows four snapshots of the apparent horizon
surfaces on the orbital plane~\cite{Alcubierre98b}. 
They show that the two black holes start well detached. 
Clearly, the `grid stretching' effect due
to the vanishing shift we used during evolution makes them appear to
grow in the coordinate space already after $8M$ of full numerical
evolution. These plots are particularly useful to extract qualitative
information about the system.  Soon after a common apparent horizon
covers the system which tends to have an increasingly close to
spherical shape at later times. This tends to set
an upper limit on the time at which linear theory can take over as has
been discussed in Ref.~\cite{Baker01a}.
It is expected that a common {\it event} horizon should
have appeared several $M$'s of time earlier.  The key physical feature
which actually makes the close limit approximation effective is that
the black holes share a common potential barrier which appears even
earlier than the common event horizon.

\begin{figure}
\begin{center}
\includegraphics[width=3.2in]{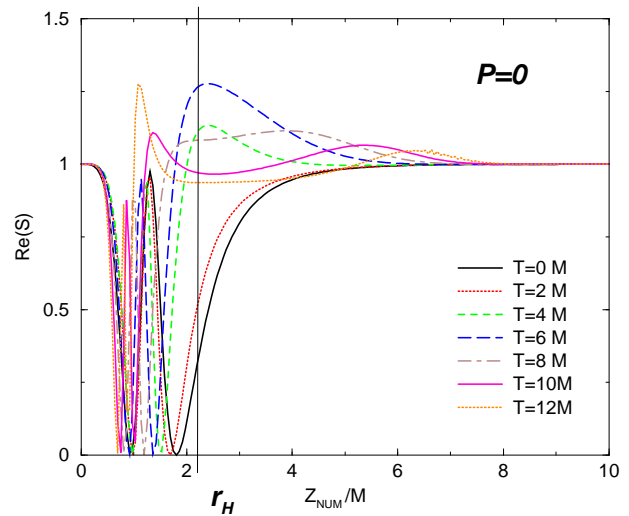}
\end{center}
\caption{S-invariant measuring deviations from Kerr}
\label{fig:rSP0}
\end{figure}

In Fig.\ \ref{fig:rSP0} we show the time evolution (actually
computed every $1M$) of the `${\cal S}$-invariant'.
The `${\cal S}$-invariant' clearly shows that space time is asymptotically
Kerr (${\cal S}=1$) at large distances. 
It also shows that the geometry at the initial slice has
significant distortions near the region where the common horizon will form
and hence we do not expect perturbation theory to be a good approximation
at very early times. In fact we need at least $4M$ of evolution to bring
the distortions outside the horizon down the level of a factor $1.5$
in the ${\cal S}$-invariant.

\begin{figure}
\begin{center}
\includegraphics[width=2.7in]{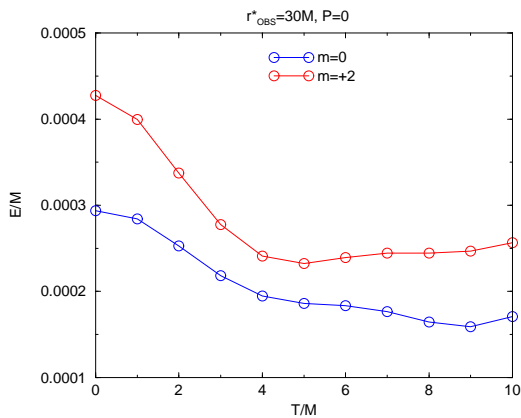}
\end{center}
\caption{Total radiated energy plots}
\label{fig:Em02P0}
\end{figure}

Fig.\ \ref{fig:Em02P0} shows the computed total radiated energy from
the system using linearized theory after $T$ of full nonlinear
numerical evolution. We see that if we insist on using perturbation
theory to compute the subsequent evolution. Even at very early times,
we obtain that the radiated energy varies notably for different
transition times. Only when we really reach the linear regime there is
a certain `energy plateau'. This again happens for $T\approx4M$ for
both relevant modes $m=2,0$. Note that the curves settle down to
values closely bearing the $1.5$ factor among mode
contributions expected for axisymmetric $l=2$ radiation at this 
orientation.

\begin{figure}
\begin{center}
\begin{tabular}{@{}lr@{}}
\includegraphics[width=2.9in]{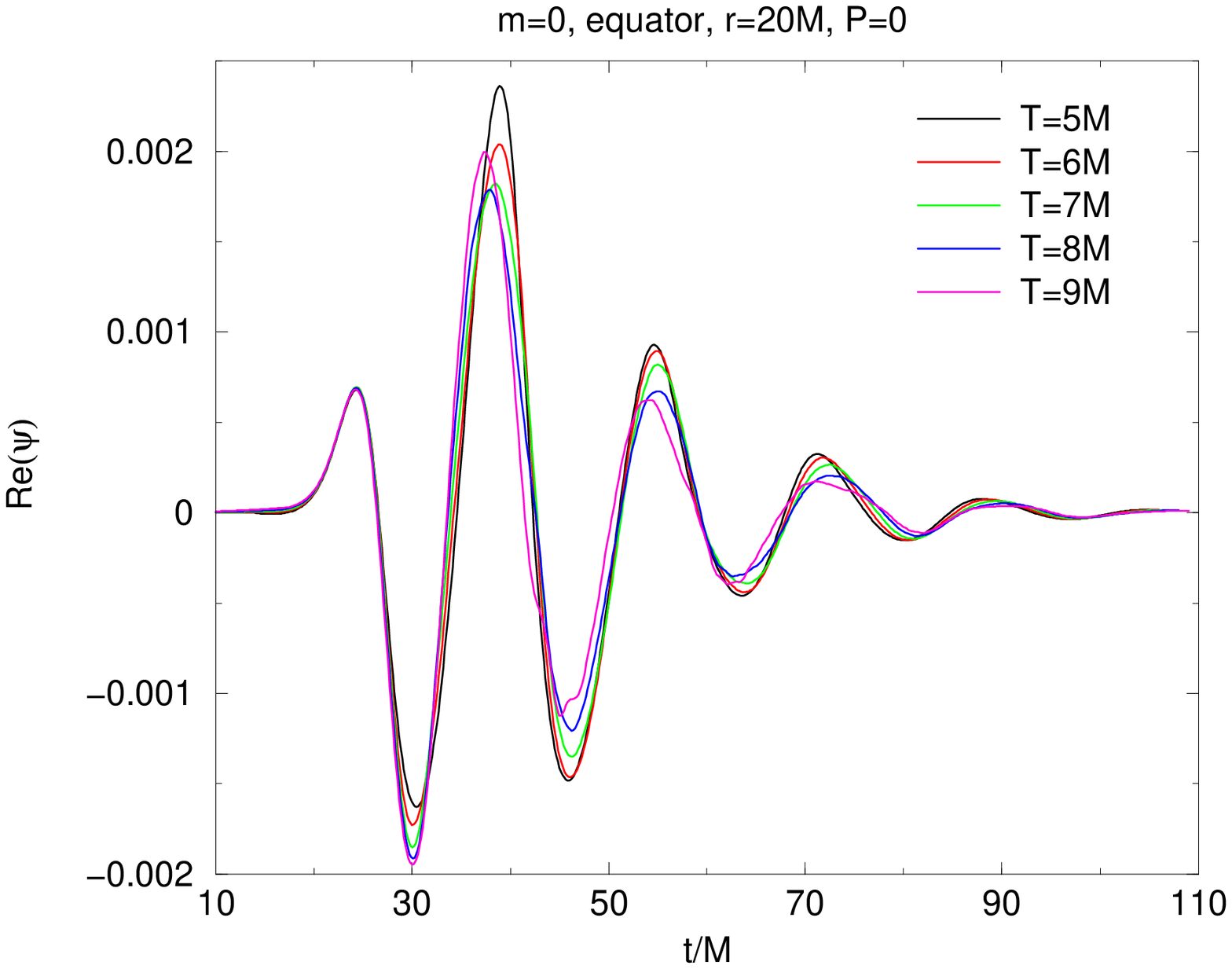}\\
\includegraphics[width=2.9in]{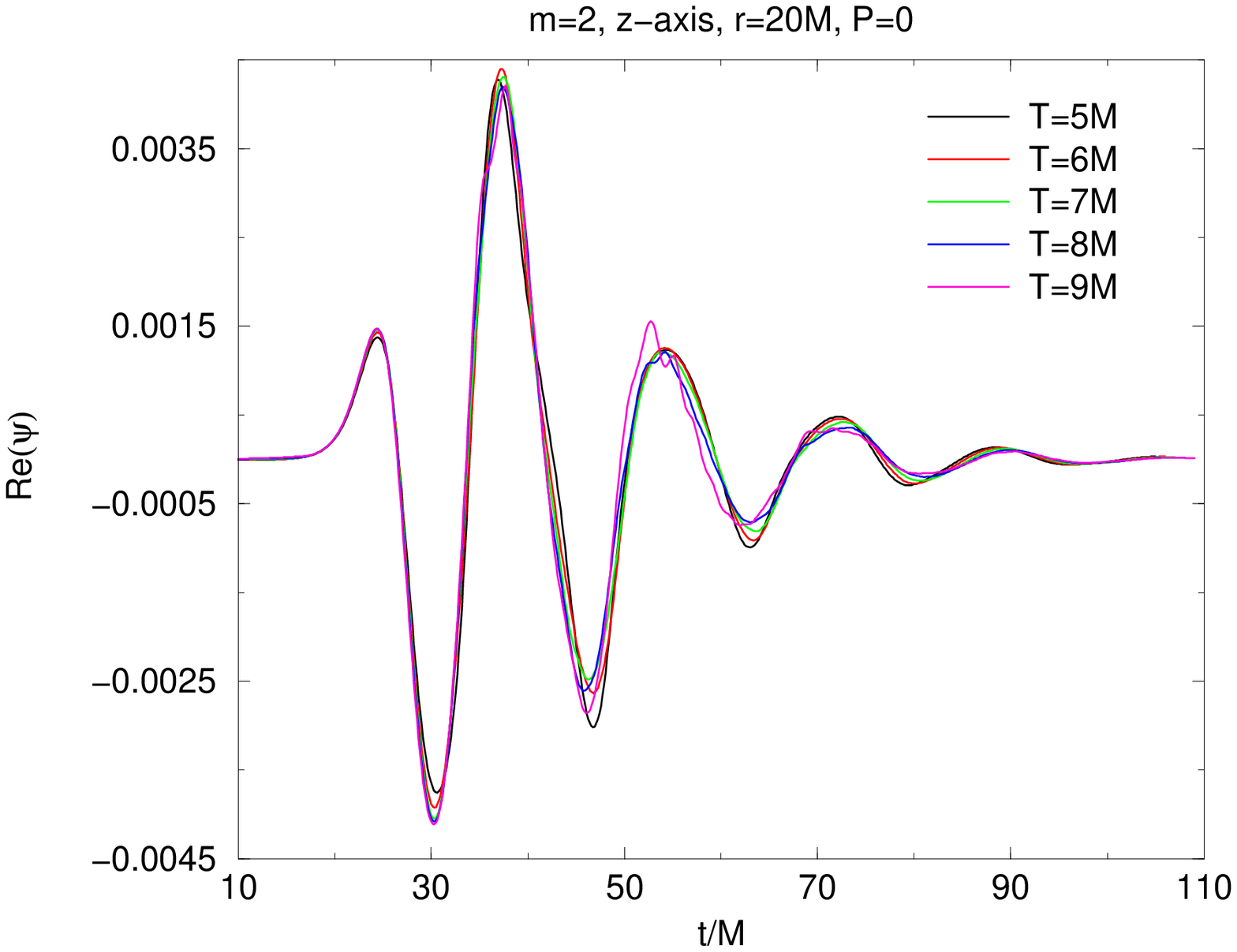}
\end{tabular}
\end{center}
\caption{Waveforms for $P_{ID}=0$}
\label{fig:WFP0}
\end{figure}

Fig.\ \ref{fig:WFP0} displays the superposition of waveforms for
the two relevant modes from $5M$ to $9M$ of nonlinear evolution.
These waveforms are unequivocally determined once the 
transition time has been chosen so there are no 
free parameters to be adjusted in this comparison. Throughout this 
section we show the $m=0$ and $m=2$ waveforms for observers respectively
on the equator and the z-axis, where the amplitude of each component is 
maximal.  
Specially useful, as an independent test of linearization, 
is the locking of the phases observed after $5M$ of
evolution. At later times we recognize the quasinormal frequency
of a the final Schwarzschild black hole with a period $\tau_{qnm}=16.8M$.

We observe
that all the four criteria used here to determine the time 
this system begins to behave linearly outside the horizon coincide
in setting it around $4-5M$ of nonlinear evolution.  Though we have 
not displayed it explicitly,
we also find good agreement with our earlier Misner data results.

\subsubsection{$P=P_{ISCO}/3$} 

Now we begin to add angular momentum to the system.
In this case we give the holes an initial momentum
perpendicular to the separation vector of one third of the
ISCO momentum magnitude. 

\begin{figure}
\begin{center}
\begin{tabular}{@{}lr@{}}
\includegraphics[width=1.8in]{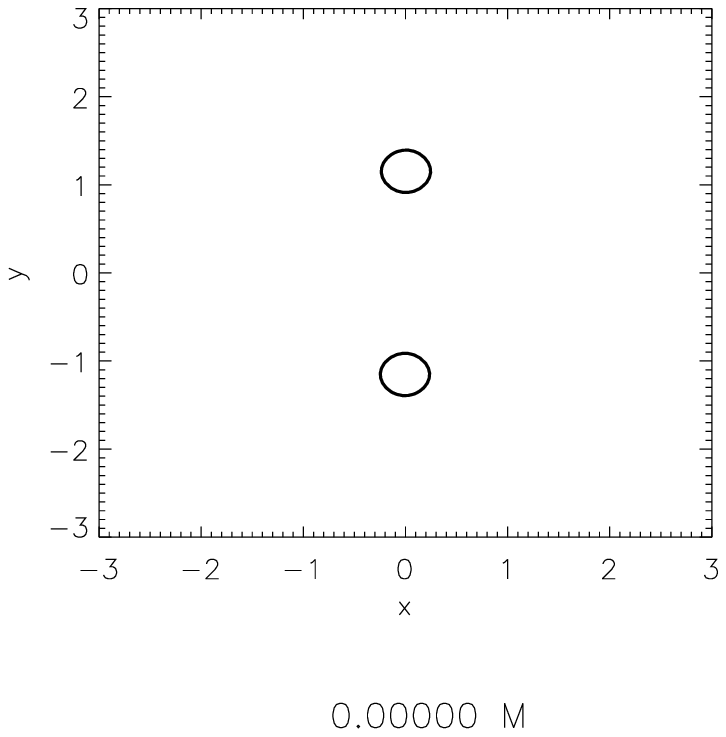} & 
\includegraphics[width=1.8in]{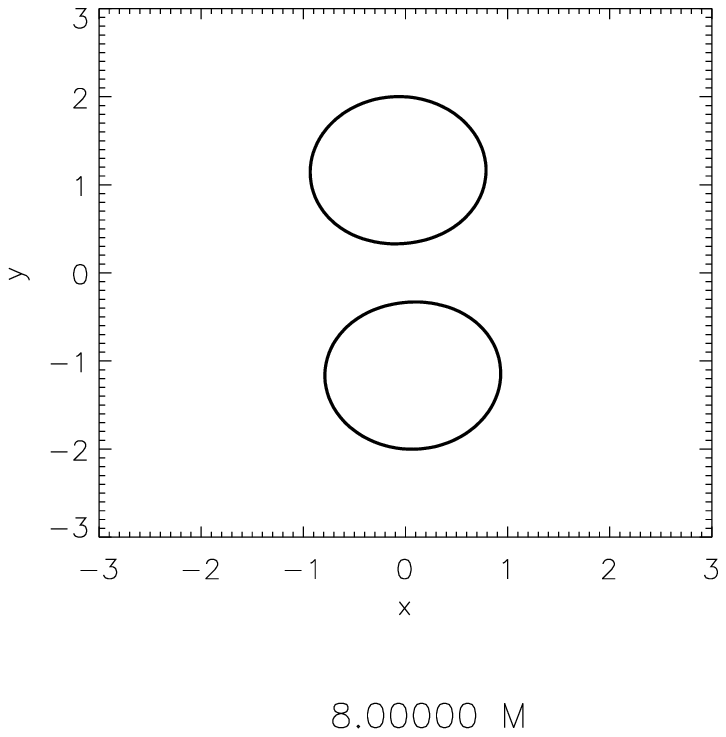}\\
\includegraphics[width=1.8in]{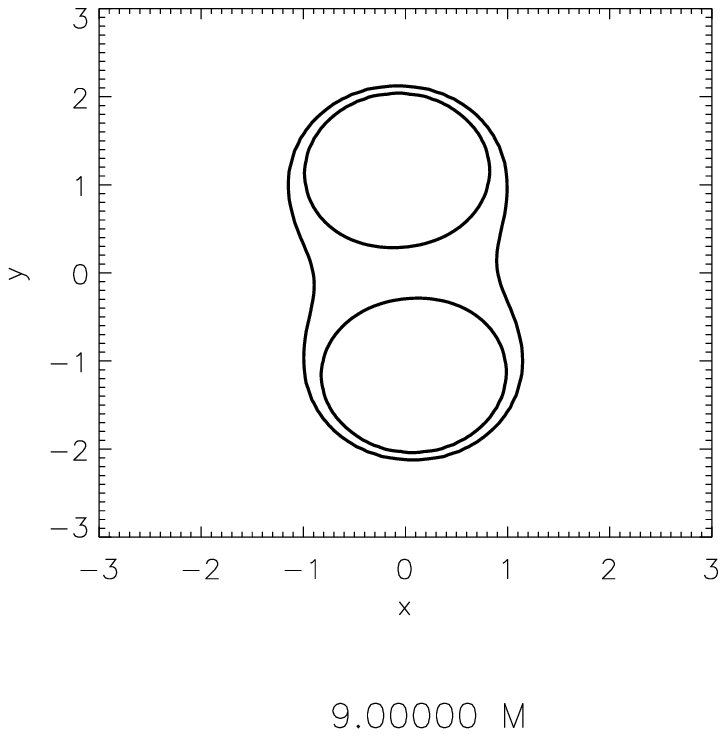} &
\includegraphics[width=1.8in]{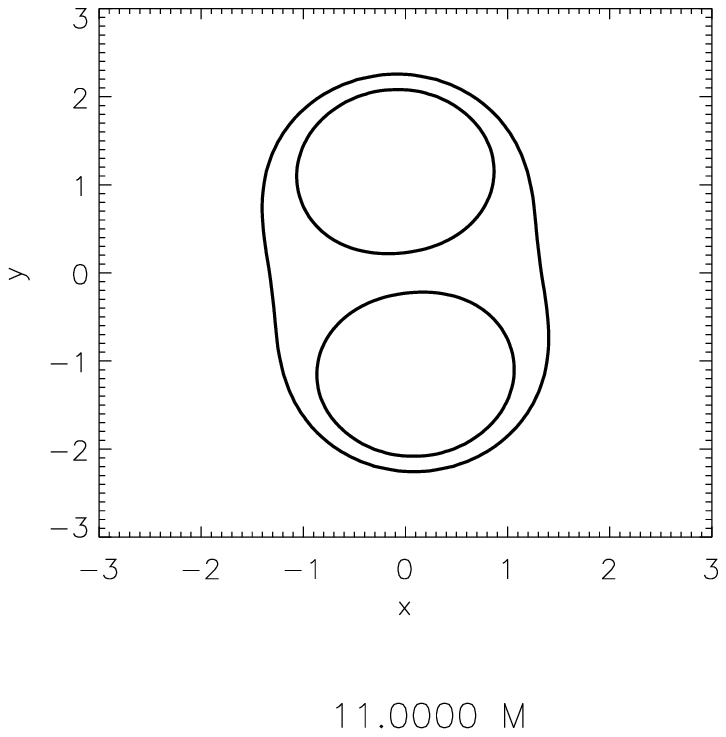}
\end{tabular}
\end{center}
\caption{Apparent horizons for $P_{ID}=P_{ISCO}/3$}
\label{fig:AHSeqP1}
\end{figure}

We show in Fig.\ \ref{fig:AHSeqP1} four snapshots of the apparent
horizon on the orbital plane; a picture quite similar to the head-on
case, except, as expected, for a slight asymmetry along the orbital
motion and the appearance of a common apparent horizon after $9M$ of
full nonlinear evolution, approximately $1M$ later than in the head-on
case.  The asymmetry is a result of the orbital-like motion with
respect to our vanishing shift coordinate system.  The coordinate
singularity (``puncture'') which must remain inside the apparent
horizon is fixed in place in the grid coordinates so that the apparent
horizon cannot ``orbit'' in these coordinates, but is, rather,
stretched around in the $\phi$-direction.  The same characteristic is
evident for all our runs. This twisting of the spacetime is a generic
consequence of using a vanishing shift for these simulation and
illustrates the importance of using a co-rotating shift with some
$\varphi-$component.

\begin{figure}
\begin{center}
\includegraphics[width=3.2in]{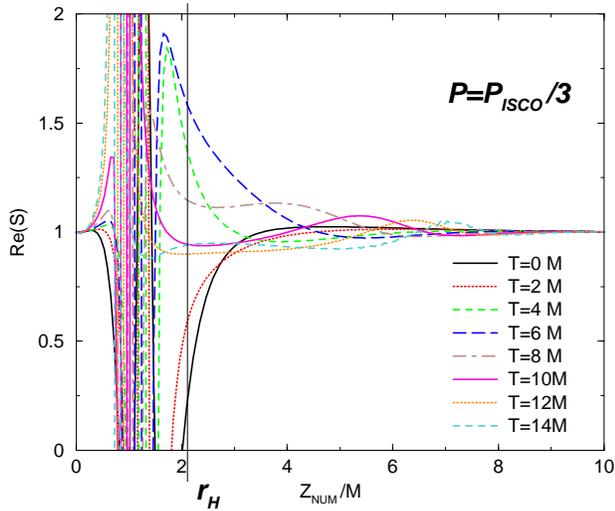}
\end{center}
\caption{S-invariant measuring deviations from Kerr}
\label{fig:rSP1}
\end{figure}

Fig.\ \ref{fig:rSP1} shows the deviations from a Kerr background along
the $z-$axis (the direction of maximal distortion) as measured by the
${\cal S}-$invariant.  The picture shows clearly that at least $4M$ of
evolution are necessary to reach a perturbative regime.

\begin{figure}
\begin{center}
\includegraphics[width=2.7in]{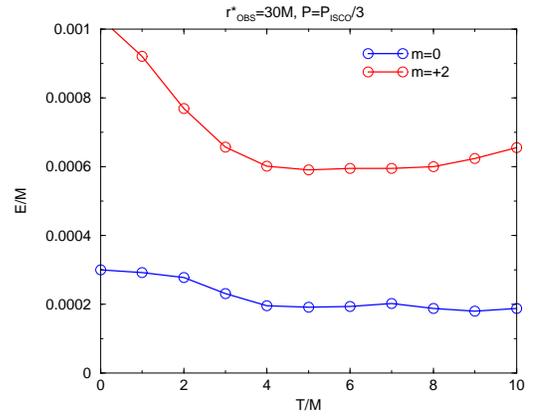}
\end{center}
\caption{Total radiated energy plots}
\label{fig:Em02P1}
\end{figure}

In Fig.\ \ref{fig:Em02P1} we observe that the radiated energy 
reaches a plateau after transition times $T\sim4M$ consistently
for both modes $m=\pm2,0$. Even for this weak radiation
the approximate constancy of the radiated energy holds up to
$T\sim10M$.

\begin{figure}
\begin{center}
\begin{tabular}{@{}lr@{}}
\includegraphics[width=2.9in]{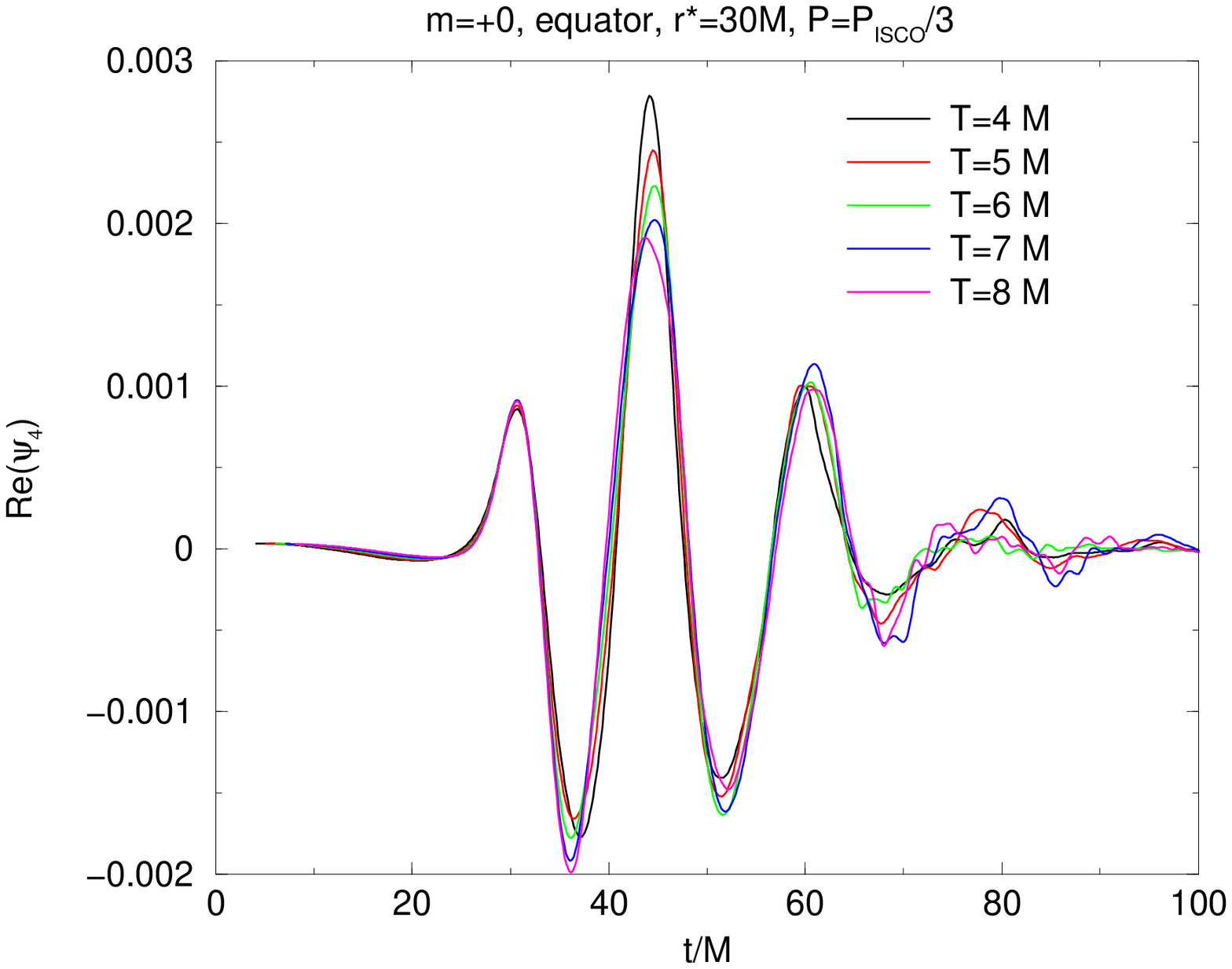}\\
\includegraphics[width=2.9in]{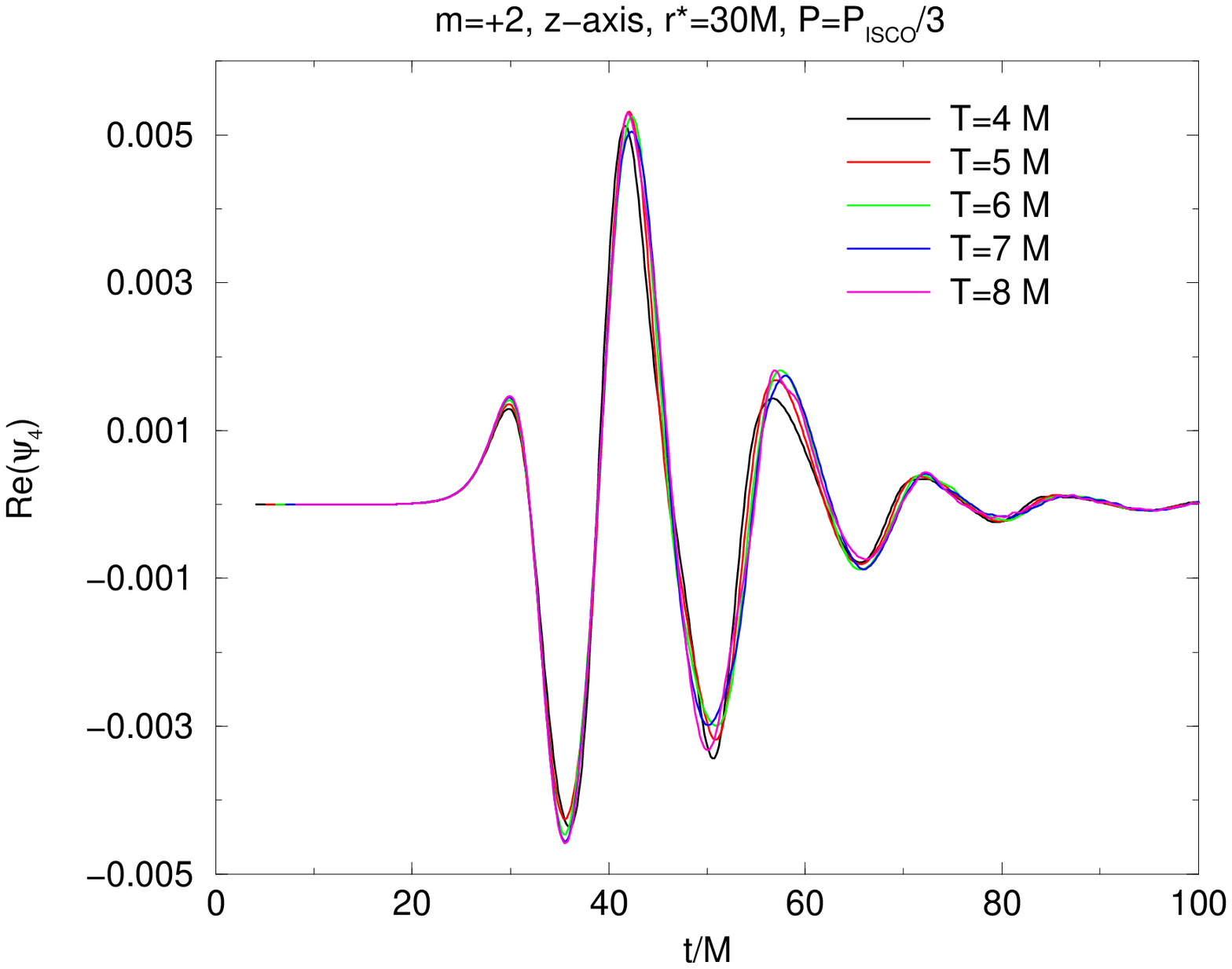}
\end{tabular}
\end{center}
\caption{Waveforms for $P_{ID}=P_{ISCO }/3$}
\label{fig:WFP1}
\end{figure}

Fig.\ \ref{fig:WFP1} displays the impressive agreement among waveforms
for different extraction times and different modes.  All this reached
without any adjustable parameter. The frequency of the waveforms quite
closely resemble those of the least damped quasinormal modes for a
Kerr black hole with rotation parameter $a/M=0.26$, which for $\ell=2,m=2$
has a period~\cite{Echeverria89} $\tau_{qnm}\approx15.2M$.

\subsubsection{$P=2P_{ISCO}/3$} 

This case has an initial transverse linear momentum of
two thirds the ISCO magnitude.

\begin{figure}
\begin{center}
\begin{tabular}{@{}lr@{}}
\includegraphics[width=1.8in]{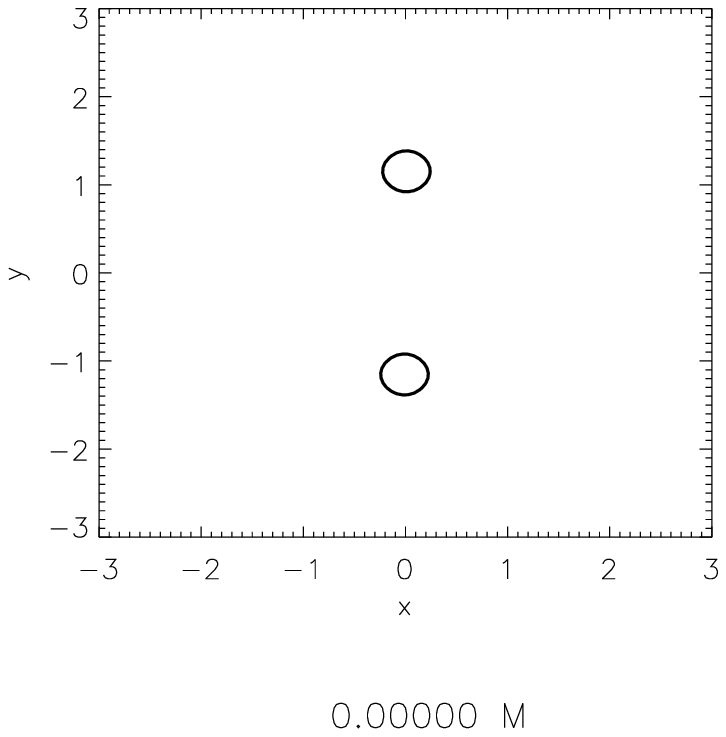} &
\includegraphics[width=1.8in]{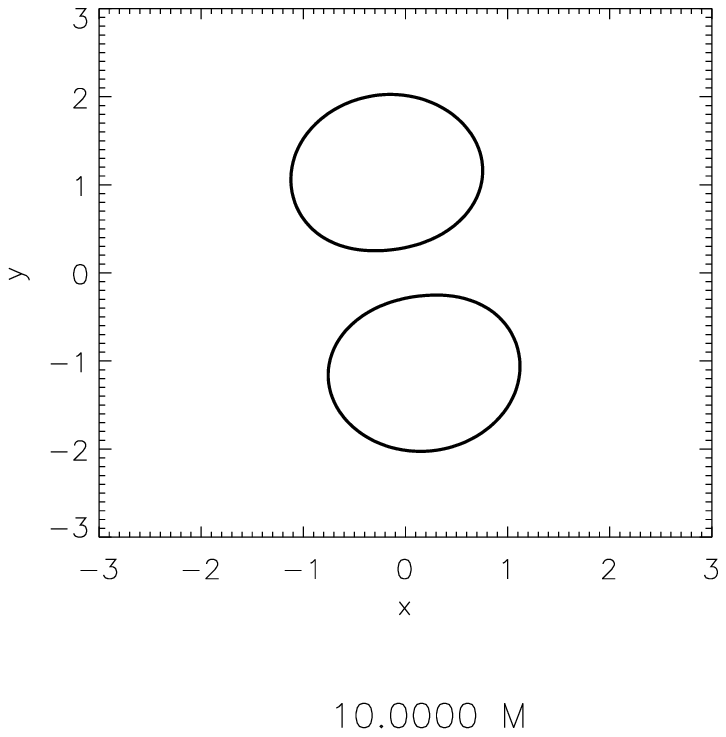}\\
\includegraphics[width=1.8in]{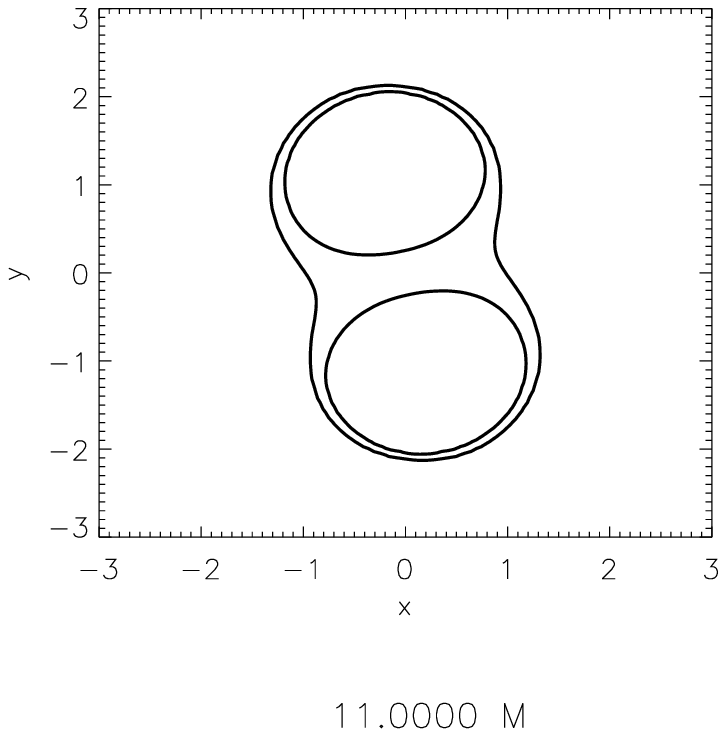} &
\includegraphics[width=1.8in]{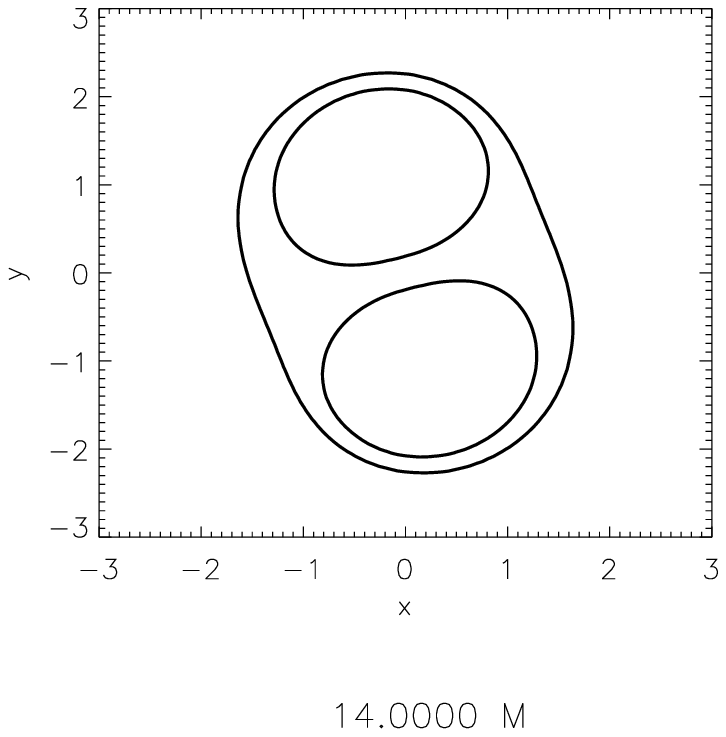}
\end{tabular}
\end{center}
\caption{Apparent horizons for $P_{ID}=2P_{ISCO}/3$}
\label{fig:AHSeqP2}
\end{figure}

Fig.~\ref{fig:AHSeqP2}
displays clear effects of orbital motion deforming
the apparent horizon surfaces on the equatorial plane.
A common apparent horizon appears at $T\sim11M$ showing
that the orbital component delays the merger of the holes
into a single one as expected.

\begin{figure}
\begin{center}
\includegraphics[width=3.2in]{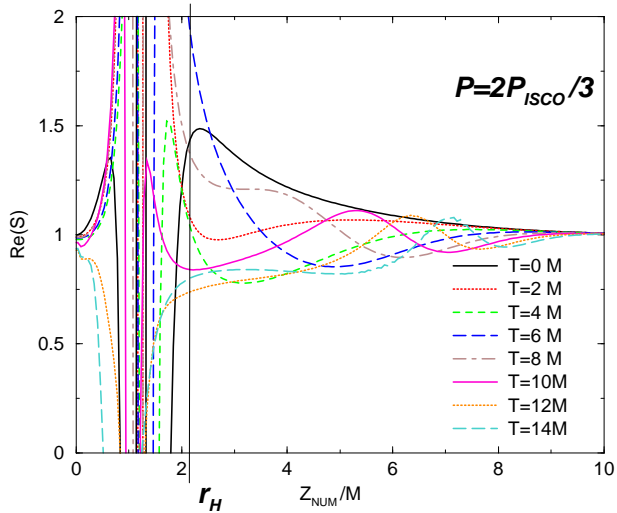}
\end{center}
\caption{S-invariant measuring deviations from Kerr}
\label{fig:rSP2}
\end{figure}

Fig.\ \ref{fig:rSP2} shows that, consistently, the ${\cal
S}-$invariant estimate for linearization is also delayed in comparison
with the head-on or near head-on cases. At least $8M$ of evolution is
needed to settle down the perturbations to an small portion of the
background.

\begin{figure}
\begin{center}
\includegraphics[width=2.7in]{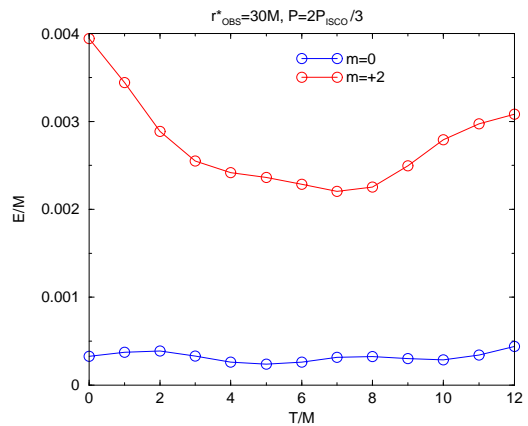}
\end{center}
\caption{Total radiated energy plots}
\label{fig:Em02P2}
\end{figure}

Likewise Fig.\ \ref{fig:Em02P2} shows how the plateau in the energy is
reached at later times after an oscillation around the final value. It
takes $9M-10M$ of evolution to obtain a stable plateau.

\begin{figure}
\begin{center}
\begin{tabular}{@{}l@{}}
\includegraphics[width=3.2in]{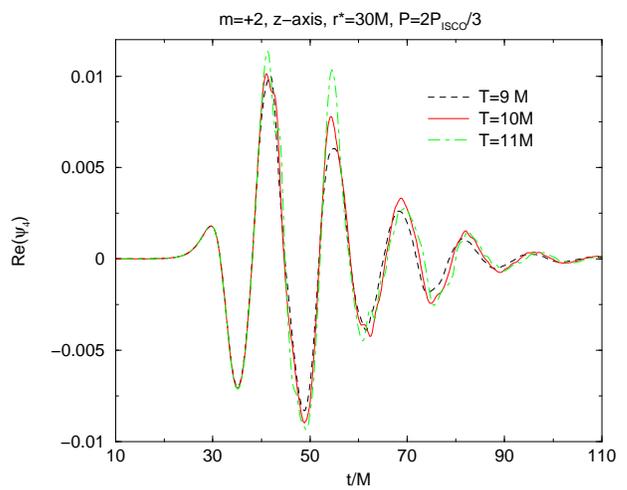}
\end{tabular}
\end{center}
\caption{Waveforms for $P_{ID}=2P_{ISCO }/3$}
\label{fig:WFP2}
\end{figure}

Fig.\ \ref{fig:WFP2} displays the leading mode, $m=+2$, along the
$z-$axis with good agreement for the times where we estimate
linearization takes place. Excellent phase locking is evident for the
first three cycles with good agreement for the rest of the relevant
signal. The frequency of the last part of the waveform agrees with the
least damped quasinormal mode of a Kerr hole with rotation parameter
$a/M=0.51$ for the mode $m=2$ which has a period~\cite{Echeverria89} of
$\tau_{qnm}\approx13M$.

For the three sets of simulations just reported we have used grids of
$384^2\times192$ size (Making use of the symmetry along the $z-$axis
of the problem) with a inhomogeneous distribution of the points
called `Fish-Eye' \cite{Baker01a}.  Each run required up to $70Gb$ of
RAM memory on a Hitachi SR-8000 at the LRZ in Garching,Germany 
and took about 10 hours running on 16 nodes (128 processors).

\subsubsection{$P=5P_{ISCO}/6$} 

To closely approach the ISCO parameters
we included the case of $P=5P_{ISCO}/6$. 

\begin{figure}
\begin{center}
\begin{tabular}{@{}lr@{}}
\includegraphics[width=1.8in]{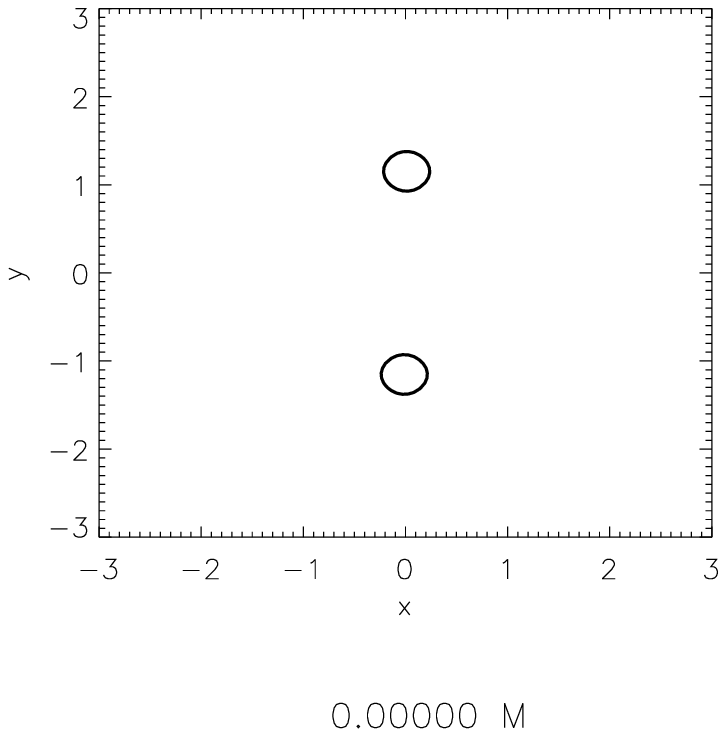} &
\includegraphics[width=1.8in]{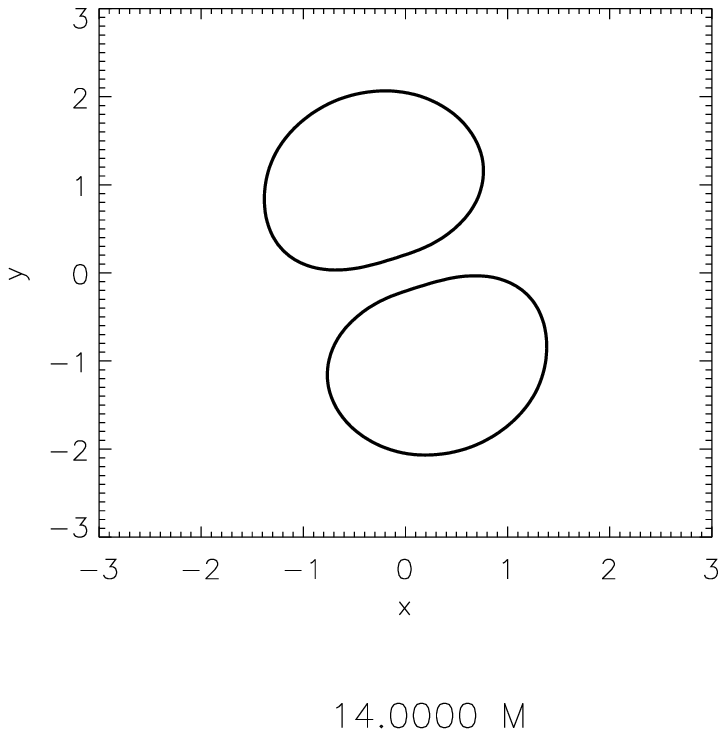}\\
\includegraphics[width=1.8in]{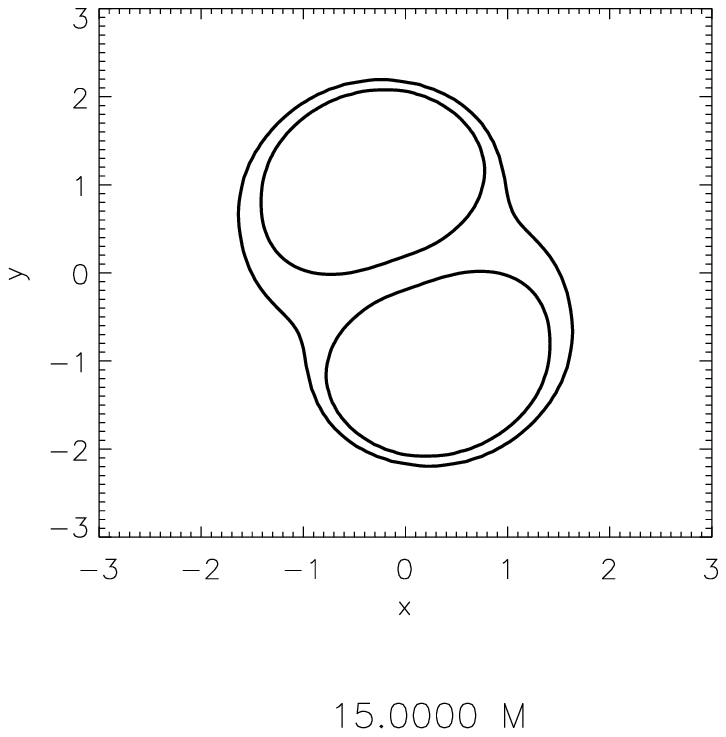} &
\includegraphics[width=1.8in]{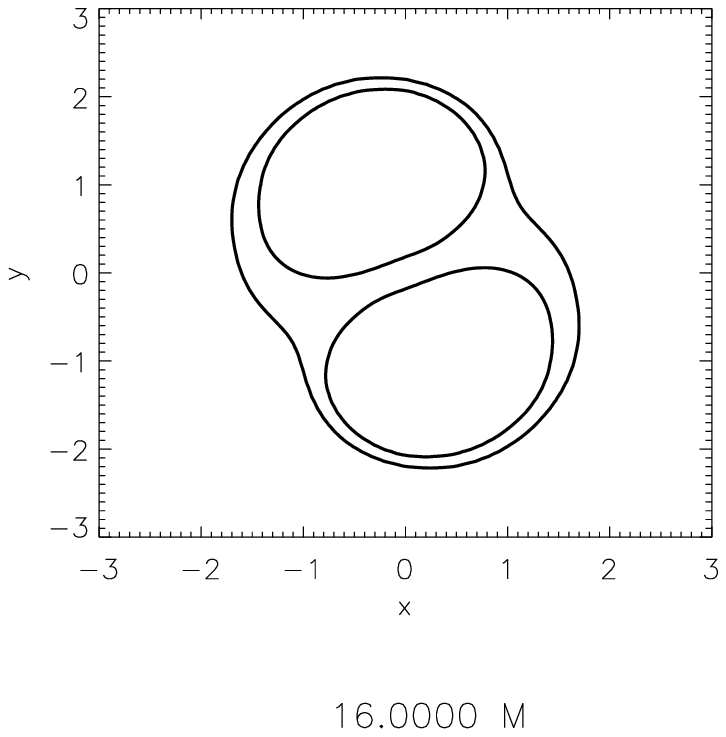}
\end{tabular}
\end{center}
\caption{Apparent horizons for $P_{ID}=5P_{ISCO}/6$}
\label{fig:AHSeqP2.5}
\end{figure}

Fig.~\ref{fig:AHSeqP2.5} shows how strongly the apparent horizon looks
deformed in the grid coordinates responding to the transverse
momentum.  The four snapshots shows the formation of a common apparent
horizon for $T_{AH}=14-15M$ of nonlinear evolution. 
\begin{figure}
\begin{center}
\includegraphics[width=3.2in]{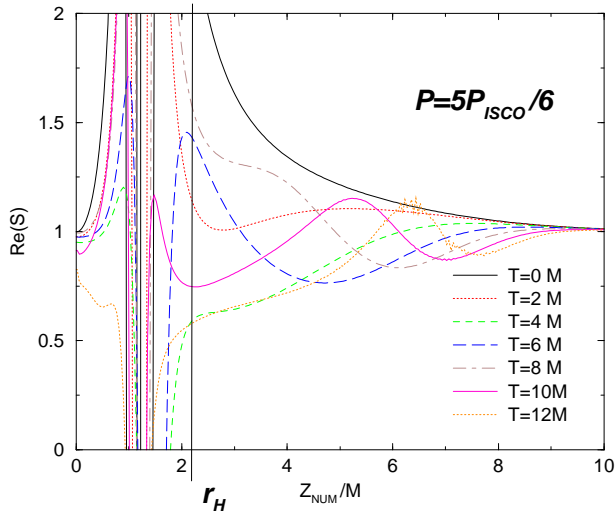}
\end{center}
\caption{S-invariant measuring deviations from Kerr}
\label{fig:rSP2.5}
\end{figure}
Consistently, 
Fig.\ \ref{fig:rSP2.5} shows the oscillations of the ${\cal S}-$invariant
around the Kerr value, $1$, and indicates that the linearization regime
is reached after about $9M$ of full numerical evolution. Note the stronger
deformation of the initial data labeled as $T=0$ compared to the previous
elements of the P-sequence.

\begin{figure}
\begin{center}
\includegraphics[width=2.7in]{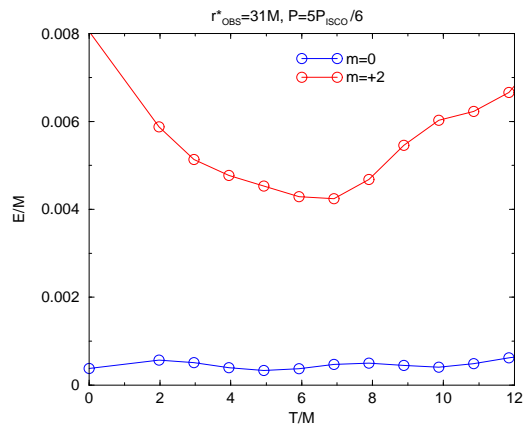}
\end{center}
\caption{Total radiated energy plots}
\label{fig:Em02P2.5}
\end{figure}

A coherent picture comes from Fig.\ \ref{fig:Em02P2.5} which 
suggests that 
energy plateau arises after $9M$ of nonlinear evolution, just before 
numerical error begins to noticeably add to the energy.. The energy
radiated oscillates around the final value and is evidently 
dominated by the modes $m=\pm2$, with very little contribution from
the mode $m=0$ and negligible from the $m=\pm4$ ones.  We note that
adding orbital momentum in moving along this P-sequence dramatically
enchances the $m=2$ component of the radiation with relatively little
effect on the $m=0$ mode.  Thus while the energy content of the two
components was roughly comparable in the head-on case, as we approach
ISCO the $m=2$ component is growing strongly as would
be expected for a non-axisymmetric orbital system.
A linear analysis~\cite{Gleiser00} indicates that the
leading $\ell=2, m=2$ contributes
to the radiated energy with a quadratic dependence in the momentum. This
is precisely what we observe for small $P$ (See Fig.~\ref{fig:SeqP}).

\begin{figure}
\begin{center}
\begin{tabular}{@{}l@{}}
\includegraphics[width=3.2in]{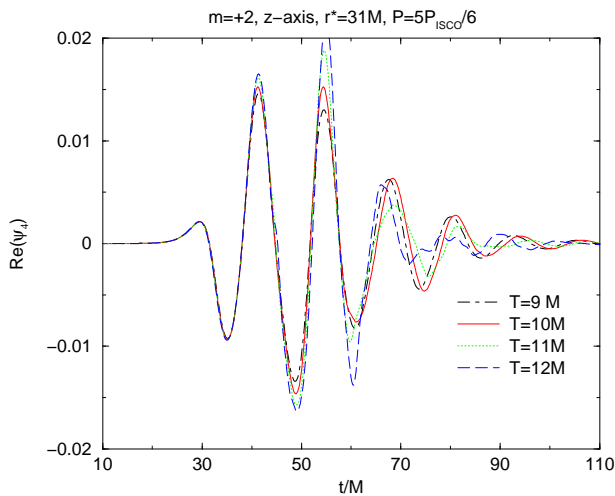}
\end{tabular}
\end{center}
\caption{Waveforms for $P_{ID}=5P_{ISCO }/6$}
\label{fig:WFP2.5}
\end{figure}

Fig.\ \ref{fig:WFP2.5} show excellent agreement of waveforms for
different transition times in the linear regime. The phase locking is
excellent up to $t\sim60M$, then there seems to be some interference
from spurious radiation from the boundaries of the full numerical
simulation that specially affects at transition times $T\geq12M$.

To compute the emitted radiation and waveforms we have made a first
correction to the Kerr background to take into account the portion of
the energy radiated $\sim 1.3\%$ and the angular momentum radiated
ending with a black hole with mass parameter
$M_{Kerr}\approx0.987M_{init}$ and $a_{Kerr}\approx0.6M_{Kerr}$.  The
normal frequency of the waveforms shown in Fig.\ \ref{fig:WFP2}
correspond quite closely to the least damped quasinormal modes of this
final Kerr hole.

This simulation and the one we will report in the next subsection
for the ISCO havei been performed on a $512^2\times256$ grid requiring
10 hours of running time on 64 nodes (512 processors) of the Hitachi
SR-8000 at the LRZ. The central resolution was $M/24$ and the
boundaries located at $37M$ where the resolution reached $\sim 1M$.

\subsection{Plunge from the ISCO data}\label{Sec:ISCO}

The ISCO configuration proves to be a borderline case to study with
the present full numerical technologies. we have chosen the standard
ADM formulation of Einstein equations in order to ensure good accuracy
during the relevant nonlinear evolution. It is known that in the ADM
formulation the instabilities that kill the evolution appear suddenly
and do not affect the early part of the numerical~\cite{Frittelli:2000uj}
integration. This is reflected in the computation of the norm of the
Hamiltonian constraint, $H$
\begin{equation}
L_2^{Norm}=\frac{\sqrt{\sum_{ijk}H_{ijk}^2}}{N_{ijk}},
\end{equation}
where $N_{ijk}$ is the total number of grid points labeled by $ijk$,
as a first measure of the numerical errors generated during the
unconstrained evolution. 

\begin{figure}
\begin{center}
\includegraphics[width=2.7in]{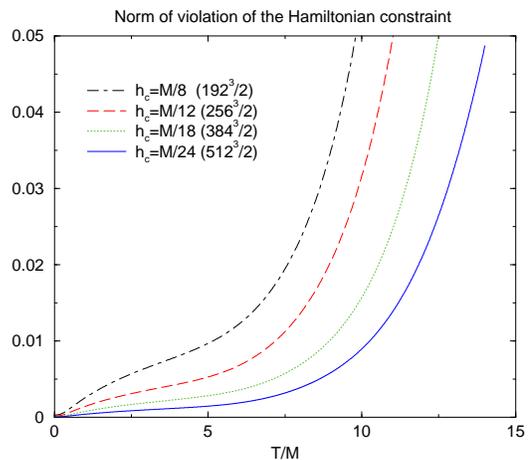}
\end{center}
\caption{Norm of the violations to the Hamiltonian constraint}
\label{fig:ham_nm2}
\end{figure}

In Fig.\ \ref{fig:ham_nm2} we show this
norm for different central resolutions $h_c$. Even though we see the
expected quadratic convergence, the exponential error growth is catastrophic
and only slightly delayed by applying significantly 
more computational resources to increase the resoultion.
Eventually these runs are killed by an non-convergent instability, 
but the numerical errors are already a problem earlier while 
the code still appears to perform convergently.
We take over with perturbation theory before the error has grown 
prohibitively.

\begin{figure}
\begin{center}
\includegraphics[width=3.2in]{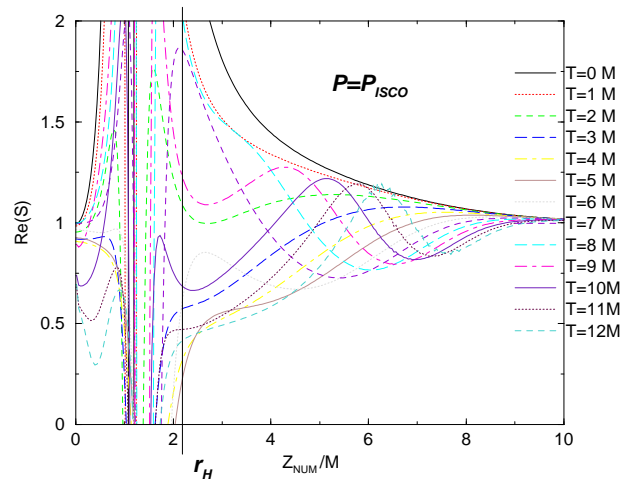}
\end{center}
\caption{S-invariant measuring deviations from Kerr}
\label{fig:rSP3}
\end{figure}

Fig.\ \ref{fig:rSP3} displays a detailed (every $1M$) evolution
of the real part of the ${\cal S}-$invariant index showing distortions
from the Kerr background (value equal $1$). Initially the distortions
are large and the subsequent oscillations maintain a significant 
amplitude near
the common horizon. After the second bounce around $T=11M$ 
the distortions begin to stabilize. At larger radii
a signal with relatively large amplitude begins to leave the system.
This will generate the first burst of radiation coming out
from the collision.

\begin{figure}
\begin{center}
\begin{tabular}{@{}lr@{}}
\includegraphics[width=2.7in]{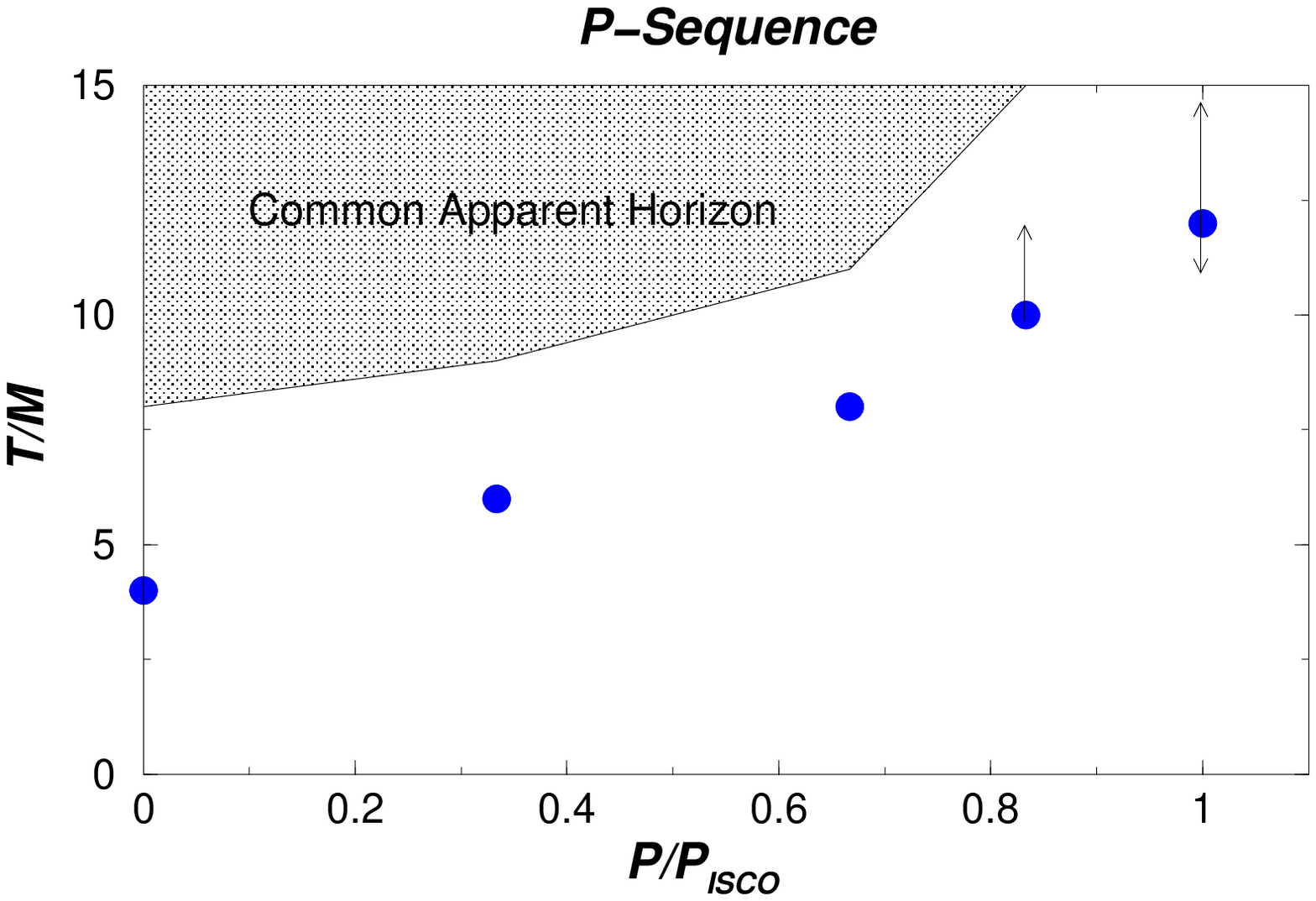}\\
\includegraphics[width=2.7in]{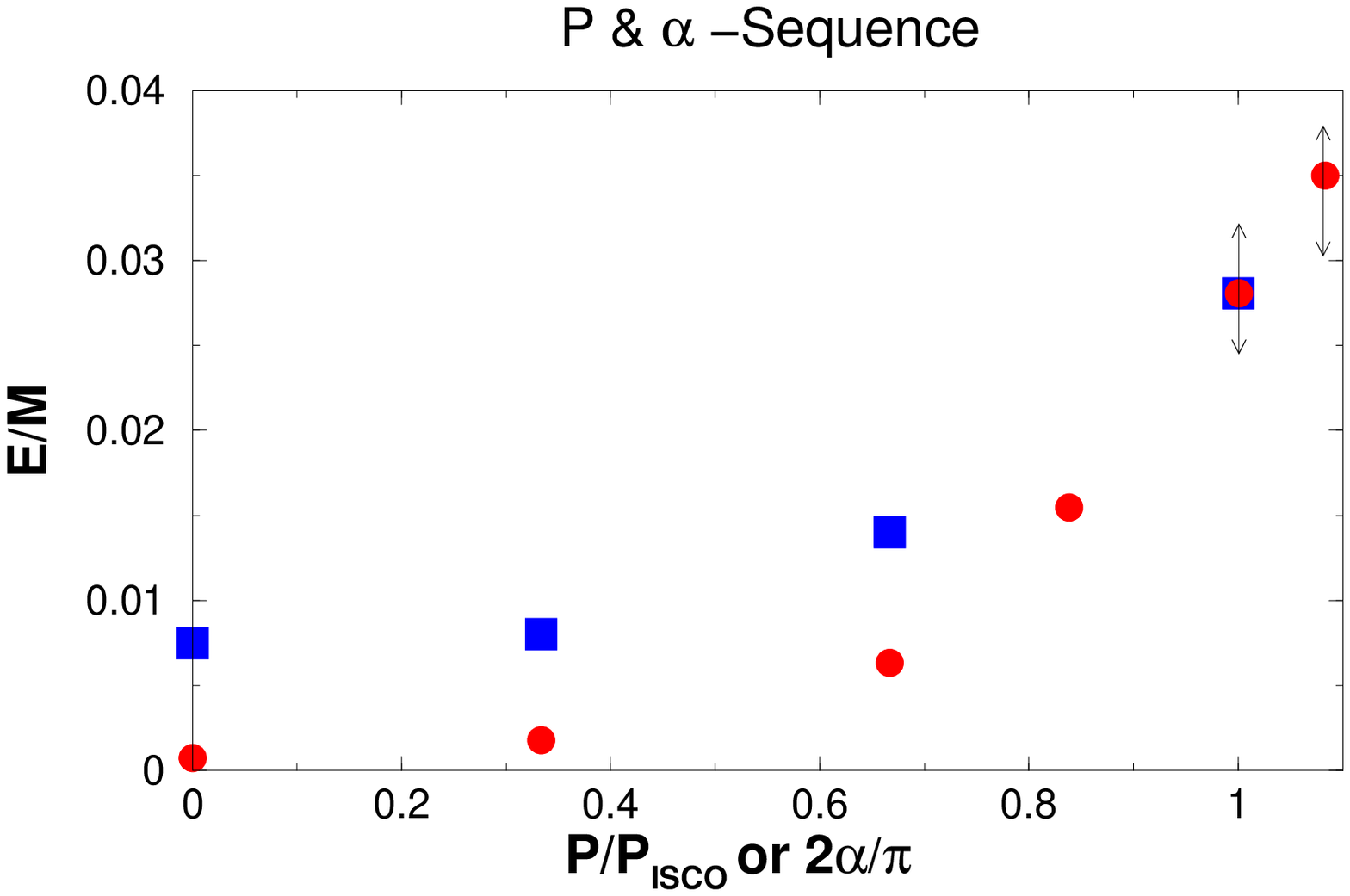}
\end{tabular}
\end{center}
\caption{Linearization time and radiated energy}
\label{fig:SeqP}
\end{figure}

In Fig.\ \ref{fig:SeqP} we gather the estimated linearization
times for the elements of the `P-sequence' and the ISCO and
compare them with the time for the formation of a common apparent
horizon. They seem to carry a similar dependence and off set by
$4-5M$ of further evolution. The offset is expected on the grounds
of relating the linearization time to the applicability of the
close limit approximations, associated to a common potential barrier,
which appears earlier in the evolution than a common event horizon and
even earlier than a common apparent horizon.

We have also considered two other sequences of 
black hole configurations in the near-ISCO regime.
The $\alpha$-sequence describes
configurations with constant separation and magnitude of the linear
momentum but varies the angle between $\vec{P}$ and the line joining the holes.
We have chosen $\alpha= 90^o, 60^o, 30^o, 0^o$ in order to consider different
radial components.
As an independent test we have also studied a sequence which connects
the ISCO to
the close limit, (where numerical simulation should not be needed), by
varying the initial separation of the holes from the ISCO separation
to $L/L_{ISCO}=1/4,1/8$.  For the sake of a compact description we are
not going to describe those results in detail apart from 
reporting the radiated energy in Fig.\ \ref{fig:SeqP}.
Note that the `$\alpha-$sequence' generates more radiation than the 
`P-sequence' as expected due to its extra radial component. In this
plot we also report on the case $P=13/12P_{ISCO}$ to check the effects
of overshooting the ISCO data, finding consistent results.

\begin{figure}
\begin{center}
\begin{tabular}{@{}lr@{}}
\includegraphics[width=2.8in]{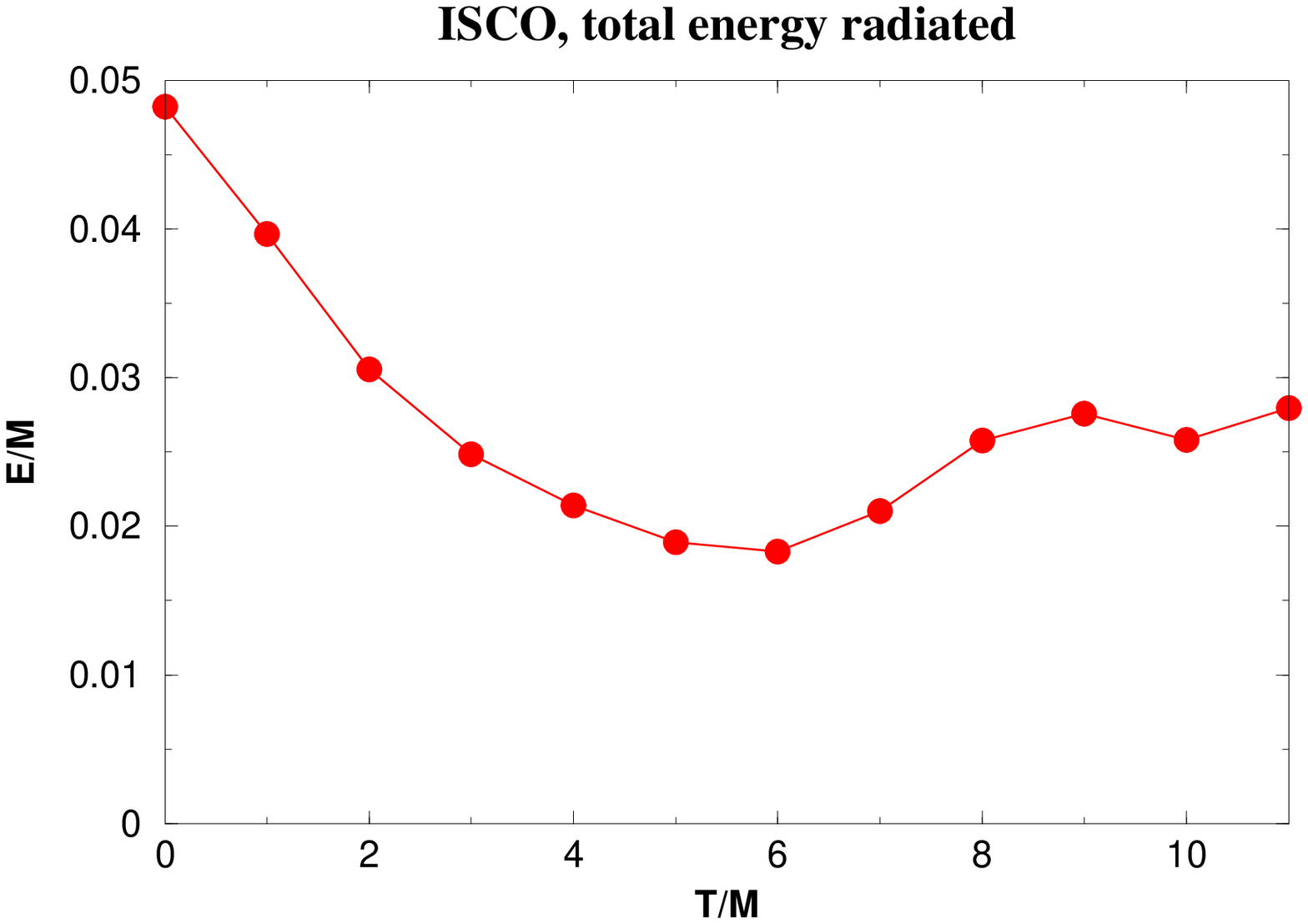}\\
\includegraphics[width=2.9in]{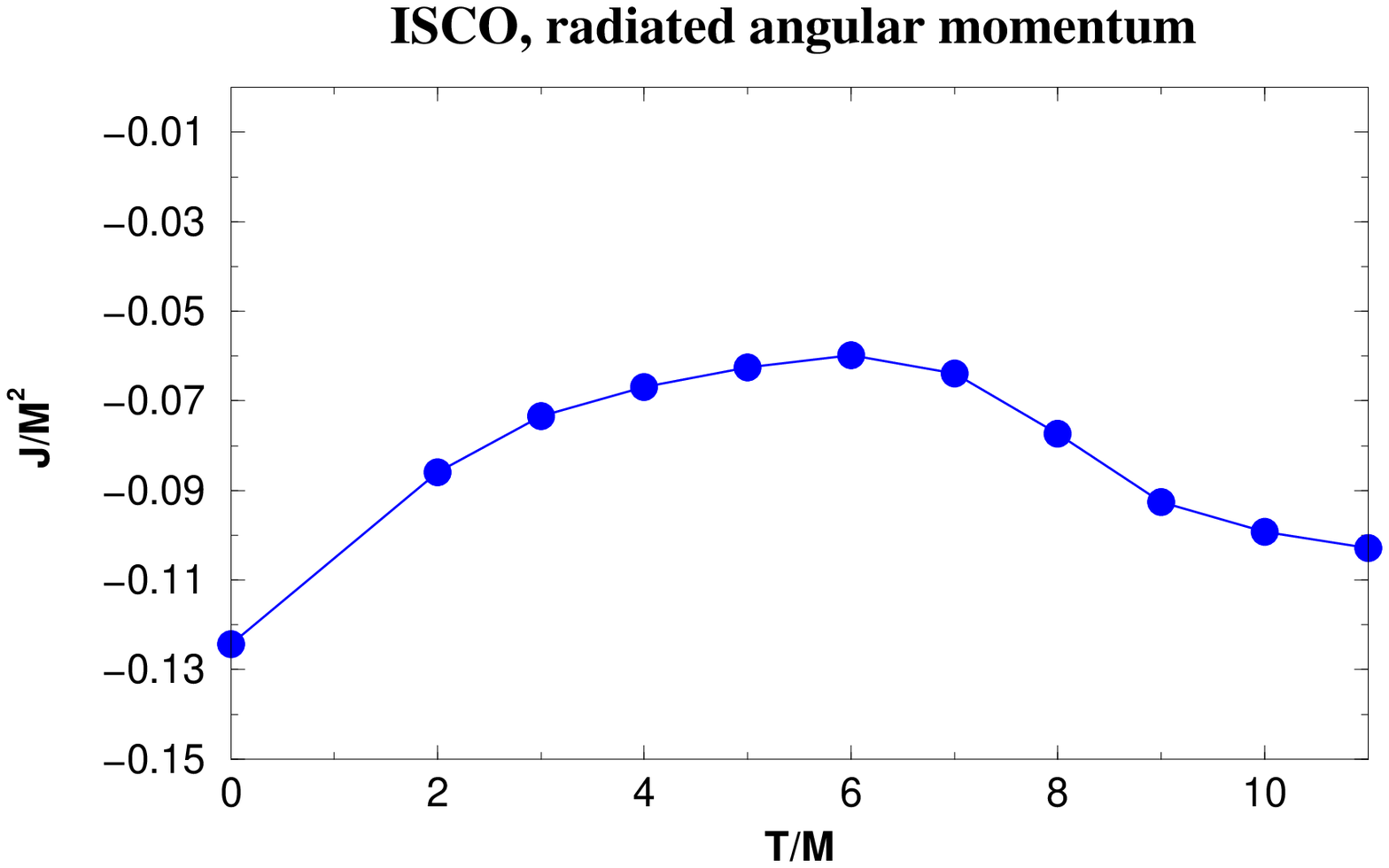}
\end{tabular}
\end{center}
\caption{Energy and angular momentum radiated}
\label{fig:EJvsT}
\end{figure}

To compute the total radiated energy and angular momentum radiated as
a function of the transition time $T$ displayed in Fig.\
\ref{fig:EJvsT}, we have taken into account a change in the background
mass $M_{Kerr}=0.97M_{initial}$ and the rotation parameter
$a_{Kerr}=0.7M_{Kerr}$.  We have also considered an extrapolation of
the results from two different resolutions in the convergent part of
the full nonlinear simulation.  In this case we find that the energy
plateau is reached a little earlier (as expected if the background is
closer to the actual one) and remains relatively constant from $T=8M$
through $T=11M$.

As remarked in Ref~\cite{Campanelli99} the computation
of the angular momentum radiated (Eq. \ \ref{angmomentum})
is a very sensitive quantity depending on correlation of waveforms. So we
can only estimate it to be around $12\%$ of the initial momentum.

\begin{figure}
\begin{center}
\includegraphics[width=2.9in]{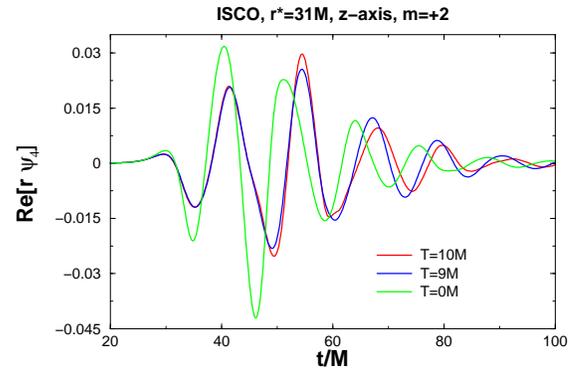}
\end{center}
\caption{ISCO Waveforms}
\label{fig:WF}
\end{figure}

The waveforms of Fig.\ \ref{fig:WF} represent our best knowledge of
the plunge so far. We have included two transition times in the quasi
linear regime to show the differences and as a measure of the internal
errors of the method. We also display the waveform obtained if no
nonlinear evolution is performed, labeled as $T=0M$. The clear
differences with the ones obtained in the linear regime show how
important is to have some nonlinear evolution that settles down the
system to a single rotating black hole plus distortions that will be
radiated away (and down onto the final black hole).

\begin{figure}
\begin{center}
\includegraphics[width=3.0in]{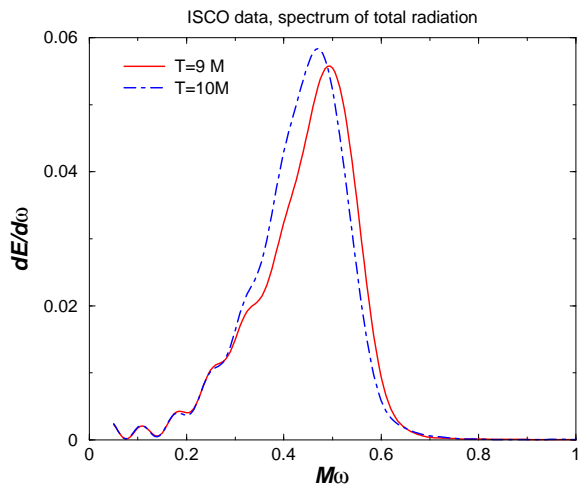}
\end{center}
\caption{ISCO spectrum}
\label{fig:spectra}
\end{figure}

An interesting feature of the waveforms studied is that they are
dominated at later times by the least damped quasinormal modes of the
final Kerr black hole. For $m=+2$ they have a period of
$\sim12M$. This is reflected in the spectrum of the total radiated
energy, including the two polarizations and integration over all
angles, presented in Fig.\ \ref{fig:spectra}.

This spectra, again shown for different transition times to have a measure of
the internal error, have a rather narrow peak at a frequency near that
of the least damped quasinormal mode for a Kerr hole with $a/M\sim0.7$.
The lower frequency part of the spectrum is dominated by the quasinormal
mode $m=-2$ component with a period of $\sim19M$ that is close to the
radiation-orbital frequency (one half the orbital period of the ISCO
$\sim37M$). For a binary system of total mass $M=35M\odot$ these frequencies
correspond to $475Hz$ and $300Hz$ respectively.

\subsection{Critical Assessment}

We have applied our late-stage model to a sequence of initial data
sets ranging from the head-on collision case to the ISCO data.  Having
achieved the expected result in the head-on case we have slowly added
orbital angular momentum.  As expected the orbital motion leads to a
dramatic increase in the leading $l=2$, $m=2$ component of the radiation,
initially depending quadratically on the momentum of the system.
  Our study proves that some fully non-linear treatment is
essential, for a reasonable treatment of the BBH system in the plunge
regime.  On the other hand we find also that the extent of the
spacetime which must be treated nonlinearly in this regime is modest.
Our estimates for the rough linearization time from our ISCO-candidate
data set come to around $10M$ of nonlinear evolution.  While the
overall results at the ISCO are only marginally robust, within say
20\% rather than much finer precision in the head-on case, many
features of the waveforms are precisely determined.  The first two
cycles or so, up to the peak at $t=50M$ or so are quite precisely
determined in phase.  In the worst case, that we have underestimated our
linearization time, and our numerical accuracy, we would not expect
dramatic changes in the early part of the waveform from an improved
calculation.  In many cases there are some evidently non-robust
features in the later parts of the waveforms.  We have traced back to
several sources of error which influence this part of the waveform,
most notably numerical differencing error originating near the
punctures in our simulations and radiation errors associated with the
finite spatial extent of our numerical domain.  This study was a vital
tool in our efforts to reduce finite boundary effects to the point
that the overall waveform seems to be accurate within 10-20\%.
Because the numerical error associated with the punctures shows up
predominately in the higher $l$ parts of the waveform, it is practical
to filter these effects by extracting the strongly dominant $l=2$
component of the radiation.  This filtering preserves some 90\%
of the $m=2$ radiation energy.  The filtered
waveforms are much more robust, and indeed simpler to characterize
(no-longer 2D).  Therefore, in the subsequent analysis of our
astrophysical waveform models we will focus on the $l=2$
radiation component.


\section{An early-stage model}\label{Sec:Model}

Two important classes of treatments are applicable to binary black
hole systems toward the end of the orbital epoch.  In addition to
`Full Numerical' approaches which attempt to solve Einstein's
nonlinear gravitational field equations are PN techniques, which
make perturbative expansions of Einstein's equations in powers of
$v/c$, and are thus most effective for describing the exterior
field of slowly moving, separated binary systems.

Having established a connection to previous studies allows us
confidence in our calculations as solutions of Einstein's equations.
But solving Einstein's equations is not sufficient for work which aims
to treat astrophysical phenomena.  For this we need a good {\em model}
of the system to be studied, and a method for evaluating the veracity
of the model. While we continue, separately, to develop an 
interface between the PN
method and numerical simulations which is ultimately expected to be
required to provide the best astrophysical initial data, such data are not yet practical.
Instead, we adopt a very basic early-stage model which takes no 
information directly from the PN treatment.  We start
our simulations with initial data derived from an alternative
description black holes in quasi-circular orbit
 derived in Refs.~\cite{Cook94,Baumgarte00a}
for conformally flat black hole initial data, which we discussed in
detail in Sec.~\ref{Sec:ISCOID}.  As there are a variety of comparable
schemes which might be applied to provide such data we also need a
means to assess the ``astrophysical appropriateness'' of this
particular scheme.  Ultimately the resolution of these issues will be
settled by performing a sequence of evolutions beginning with more and
more separated initial systems approaching the regime where different
descriptions of the initial configurations (Post-Newtonian,
quasistationary relativistic ansatz, etc) merge together.

In Sec.~\ref{Sec:QCseq} we describe such a sequence and in
Sec.~\ref{Sec:PNparameters} compare our results with PN calculations
applicable for well-separated systems.  The comparison seems to
justify our initial model as a plausible {\em, a priori} agreeing with
PN analysis to the level of accuracy presently available.  The course
for future research is to refine these calculations pursuing ever
improved initial models beginning the simulation earlier and earlier
in the astrophysical inspiral orbital process, thereby evaluating and
improving the astrophysical accuracy of these early results.  In
Sec. \ref{Sec:QCseq} we take the first few steps along this course.

\subsection{The QC sequence}\label{Sec:QCseq}

\begin{figure}
\begin{center}
\includegraphics[width=3.2in]{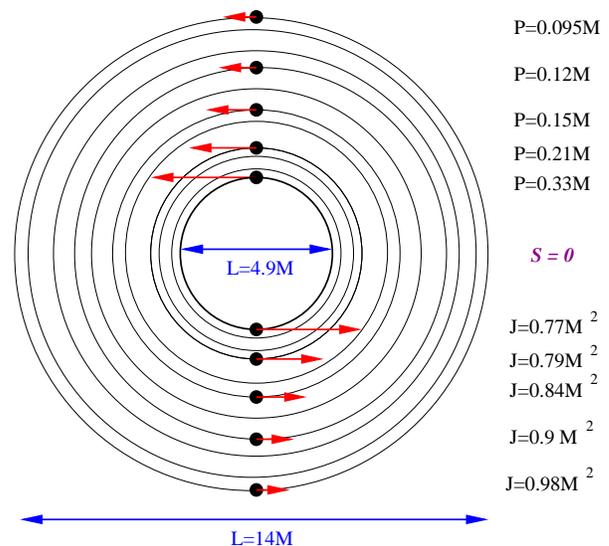}
\end{center}
\caption{Quasi-circular orbital data}
\label{fig:preISCO}
\end{figure}

Our ``early-stage'' model is composed essentially of a sequence of
spacetime slices (initial data sets).  We take these from the same
family of solutions of the initial value equations which we have used
in Sec.~\ref{Sec:latestagemodel}, ''puncture-data'', and select data
corresponding approximately to black holes in circular orbit. We adapt
Cook's results, applying the effective-potential method to non-spinning black
holes\cite{Cook94}, to our case to identify a sequence of
quasicircular orbits. This selection, which we call the ``QC''
sequence, includes data ranging in initial separation from $5M$, the
ISCO value, to $14M$.  Strictly speaking Cook's results apply to
a family of data determined with a different interior boundary 
condition than applied for our puncture-data, 
but the practical differences are very small.
Baumgarte\cite{Baumgarte00a} has performed an effective potential
analysis for the puncture data showing results almost identical to
Cook, consistent with the noted numerical similarity of the two
treatments \cite{Abrahams95c}.  Though not ideal, it is not true that
these conformally flat initial data should completely fail to
approximate an astrophysical configuration of black holes.  For
sufficiently well-separated binaries of non-spinning black holes the
uncertainties associated with this method can be made arbitrarily
small, and comparisons with post-Newtonian calculations have shown a
reasonable correspondence in this limit.\cite{Cook94} The black hole
configurations associated with our QC sequence are illustrated in
Fig.~\ref{fig:preISCO}.  

Other families of initial data are available 
which might be applied for an alternative, similarly constructed 
'QC-sequence', but none are as yet sufficiently developed for this 
application, and there is no clear way to evaluate the the various
{\em a priori} preferences adopted in these various methods without 
a dynamical study, such as that of Sec.\ref{Sec:QCresults}.
With our simple model we make no attempt to
contain information from the previous dynamics of the spacetime 
({\em i.e} radiation), but attempt a fair representation of the system's 
bulk motion at late times. 

We have converted the QC data into actual physical parameters to be
used as the initial input of the full numerical simulations.  In
Table~\ref{QCtable} $L$ is the proper distance between the throats of
the holes, $X$ is the coordinates of the `punctures' in the conformal
space, $P$ is the individual linear momentum of the holes, $J$ the
total angular momentum of the configuration, $\Omega$ the angular
velocity as measured at infinity, and $m$ is the bare mass of the
punctures. All quantities are normalized to the total ADM mass of the
system $M$.  Baumgarte's determination, QC-B, of the ISCO explicitly 
within the puncture-data family is also shown for comparison.

\begin{table}
\caption{Quasi-Circular data}
\begin{ruledtabular}
\begin{tabular}{lllllll}\label{QCtable}
Name&$L/M$&$\pm X/M$&$\pm P/M$&$J/M^2$&$M\Omega$ &$m/M$\\
\hline
QC-B \protect\cite{Baumgarte00a}&
4.90      &    1.158 &   0.334 &   0.773 &   0.176 & 0.450\\
QC-0 \protect\cite{Cook94}&
4.99      &     1.169 &   0.333 &   0.779  & 0.168 & 0.453 \\
QC-1 &
5.49      &	1.364   & 0.286  &  0.781  & 0.142 & 0.463\\
QC-2 &
5.86      &    1.516  &  0.258  &   0.784  & 0.127  & 0.470\\
QC-3 &
6.67      &   1.849 &   0.2148 &   0.794 & 0.1019  & 0.477\\
QC-4 &
7.84      &   2.343  &  0.1743 &   0.817 & 0.0760 & 0.483\\
QC-5 &
8.84      &  2.772  &  0.1514  &   0.8397 & 0.0612 & 0.487\\
QC-6 &
9.95      &  3.257  &  0.1332 &   0.8677 & 0.0497 & 0.489\\
QC-7 &
11.11      &  3.776  &  0.1190 &   0.8985  & 0.0408 & 0.492\\
QC-8 &
12.17      &  4.251  &  0.1091 &   0.9270 & 0.0348 & 0.494\\
QC-9 &
13.31      &   4.77 &   0.1005 &   0.9584 & 0.0297 & 0.496\\
QC-10 &
14.22      &   5.19  &  0.0947 &   0.9826 & 0.0267 & 0.498
\end{tabular}
\end{ruledtabular}
\end{table}

\subsection{Correspondence with Post-Newtonian parameters}
\label{Sec:PNparameters}

In Ref. \cite{Buonanno00a} Buonanno and Damour (BD) describe how to
how to use their conservative 2PN order
Hamiltonian\cite{Buonanno:1998gg} to estimate the location of the last
stable orbit of a binary black hole system.  In that paper they go on
to evolve such systems at 2.5PN order producing an estimate for the
gravitational radiation but here we are only considering the
{\em conservative} system.  Going beyond the 2PN order treatment, Damour,
Jaronowski and Sch\"afer (DJS) have resummed the conservative part of
the 3PN Hamilonian to provide a higher-order evaluation of
ISCO~\cite{Damour:2000we}. Until recently, the description of 3PN
dynamics has been dependent on an unresolved regularization ambiguity
$\omega_s$, but a recent dimensional regularization
treatment\cite{Damour:2001bu} seems to fix this ambiguity with the
result, $\omega_s=0$.  Accepting this result we can apply the methods
in \cite{Damour:2000we} to produce a sequence of quasi-circular orbits
comparable to our QC sequence.  There are two approaches to PN
resummations recommended in Ref.~\cite{Damour:2000we}.  Leaving the
reader to find the details in the above references we note only that
one approach, called the $j^2-method$ is modeled after the
resummation approach of Damour-Iyer-Sathyprakash \cite{Damour_T:98}
and the other is an extension similar to the Buonanno-Damour treatment
described above.  The cleanest way to compare the results of these
analysis is by looking at the dependence of the angular momentum $J$
on the orbital frequency $\omega$, a gauge invariant comparison.  In
this view the minimal value in the $J$ curves indicates the ISCO.
Following DJS we have also provided the results of these treatments at
the 1PN and 2PN orders

\begin{figure}
\begin{center}
\vskip 0.2in
\includegraphics[width=3.2in]{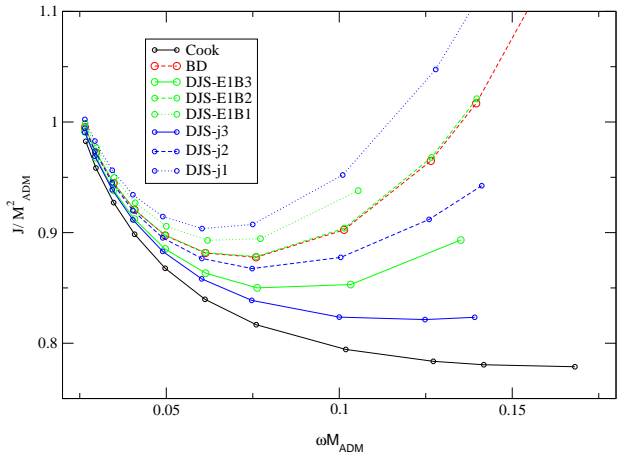}
\end{center}
\caption{The dependence of angular momentum on orbital frequency 
for several approaches.  The results indicate that the QC sequence
and ISCO is entirely consistent with the PN treatment at the level
of the precision presently possible.}
\label{fig:Jvomega}
\end{figure}

Fig. \ref{fig:Jvomega} provides a summary of several results. In this plot
we can assess several important considerations.  The reliability of
the PN treatment is accessible in two ways.  The degree of convergence
with higher PN order provides some indication of how much these
results can be trusted.  Earlier hopes that the one-body resummation
approach might provide accurate waveform information up to and even
beyond ISCO, possibly minimizing the need for difficult numerical
treatments are not supported.  The one-body approach is not
converging, in the sense that the 2PN curve is everywhere closer to
the 1PN curve than to the 3PN results.  The $j^2-method$ seems to show
slow convergence for well-separated cases with small values of
$\omega$ say $0.05$ or less.  Likewise, there is notable sensitivity
to the choice of resummation method for $\omega<0.06$ or so in the 2PN
and 3PN cases.  For the 1PN case, the one-body treatment is much
closer to the higher order treatments than the 1PN results.  It seems that
the careful crafting of the one-body resummation method has managed to
minimize the significance of 2PN terms for close agreement between the
1PN and 2PN results, but this treatment has not been done as much for
the heretofore unknown dependence on the higher PN order terms, thus
producing the appearance of anti-convergence in the one-body resummed
PN sequence.

Comparing the post-Newtonian sequences with our QC sequence it is
notable that the PN results uniformly approach the QC sequence with
increasing PN order.  Surely the limit of the PN sequence curves
approaches something other than the QC curve, but no difference
between our curve and the apparent limit of the PN sequences is yet
discernible at 3PN order.  Consistently the location of ISCO estimated
by the PN treatments approaches the parameters of the effective
potential ISCO data we used to the level of precision at 3PN.  
This analysis suggests that the parameters of our ISCO are at least 
as dependable as those given at 3PN and in fact entirely consistent 
with the PN sequence to the degree of precision currently possible.  
The PN results suggest that $\omega_{ISCO}\geq 0.075$ but do not provide 
a clear upper bound.

\begin{figure}
\begin{center}
\vskip 0.3in
\includegraphics[width=3.2in]{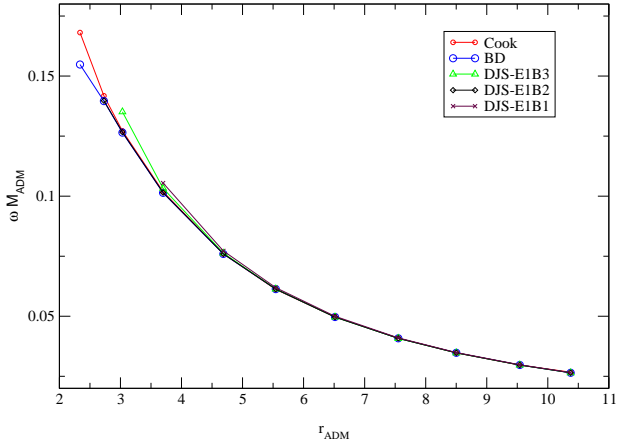}
\end{center}
\caption{The dependence of orbital frequency on ADM-gauge separation shows 
remarkable agreement among all cases.}
\label{fig:omegavr}
\end{figure}

\begin{figure}
\begin{center}
\vskip 0.2in
\includegraphics[width=3.2in]{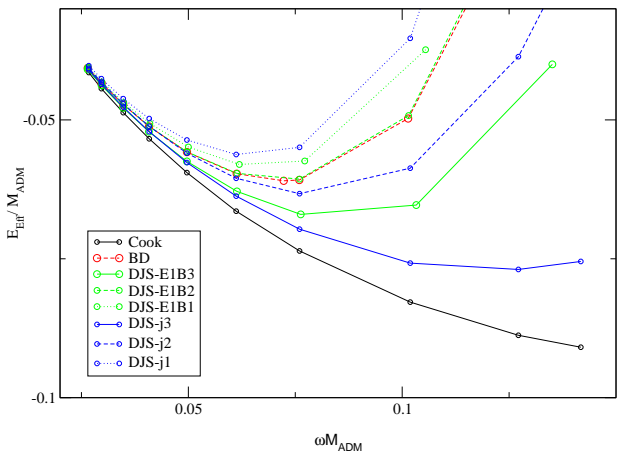}
\end{center}
\caption{The dependence of the
effective potential $E_Eff$ on orbital frequency 
for several approaches is consistent with the results for angular
momentum shown in Fig.~\protect{\ref{fig:Jvomega}}}
\label{fig:Evomega}
\end{figure}

Going beyond the comparison of the invariants $J$ and $\omega$, we can
make comparisons in gauge.  The ADM gauge applied in the one-body
approach is a directly generalization of the isotropic coordinates
applicable to conformally flat data.  We can thus compare the
separation dependence directly for the one-body treatments and QC
sequence.  (The gauge transformation for the $j^2$-method is
impractical.)  The comparison in Fig. \ref{fig:omegavr} shows profound
agreement among all treatments all orders for separations $r_{ADM}>4
M_{ADM}$. For closer separation the PN gauge transformation begins to
fail (cannot be solved) in some cases.  The other relevant quantity is
the effective potential $E=(1-M_{rest}/M_{ADM})/\nu$ shown in
Fig. \ref{fig:Evomega}.  The results are consistent with the
conclusions from the $J$-$\omega$ comparison.

It is worth remarking here that the above agreement is also 
consistent with some recent results using the standard 3PN 
expansion in the standard Taylor form, without using resummation 
tecniques \cite{Blanchet:2001id}. 

\section{Analyzing and testing the full model}\label{Sec:QCresults}

A fundamental feature of the Lazarus approach for binary black hole
spacetimes is
that have simultaneously at hand alternative approaches applicable to
the interface regions of our model spacetime.  We have extensively
demonstrated how we can take advantage of this feature at the FN-CL
interface to cross-check the results of the CL and FN treatments
against each other, and indeed to better understand the physics of the
spacetime.  Even with our very simple model for the early-part of the
spacetime we can begin applying the same type of test at the FL-FN
interface.

\begin{figure}
\begin{center}
\includegraphics[width=2.9in]{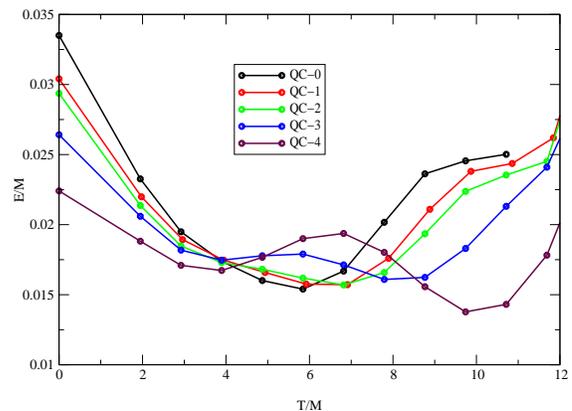}
\end{center}
\caption{Radiated energy for the QC-sequence.}
\label{fig:QCEvsT}
\end{figure}

The principle is the same, now, instead of altering the time of
transition from FN to CL we alter the time of transition from FL to
FN, and compare the resulting waveforms.  In practice this means we
select alternative data sets from the QC sequence and compare the
waveforms.  If our model is astrophysically reliable, then the shape 
of our waveforms should be independent of variations in the choice
of QC-sequence data set (labeled by a QC-number).
In this section we will compare the waveforms generated over a 
broad range of the QC sequence, treating the cases ranging from QC-0 
to QC-4 with initial proper separations ranging from about 5.0M to 
about 7.8M.
In each case we find a total radiation energy of about 2.5-3\%M during 
the course of the burst.  As described in Sec.~\ref{Sec:methods}, we
look for a plateau in the dependence of the total radiated energy on the
FN-CL transition time T, shown here in Fig.~\ref{fig:QCEvsT}.
The several methods described in
Sec.~\ref{Sec:methods} collectively suggest linearization time of around 
$T\sim9-10,10-11,10-11,11-12,13-14M$ respectively for case QC-0 through QC-4.
In all cases numerical inaccuracies begin to affect our results 
in the vicinity of 
$T=12-13M$ as is apparent in the figure so the later case, especially QC-4 
are only very marginally linearized before numerical problems are significant.
As is discussed in more detail below, we find that significant amounts of 
angular momentum are radiated away. For the lower separation cases, 
QC-0 -- QC-2 roughly $.10-.11M^2$ of the angular momentum is lost, enough to 
significantly alter the value of the close-limit background black hole
spin parameter $a$, so for the results presented in this section we
have set the 
background black hole angular momentum to the value given in
Table \ref{QCtable}\ {\em minus}\ $0.11M^2$.
\footnote{Note that in Ref.\ \protect{\cite{Baker01b}}
we have underestimated the angular momentum radiated due to a mistake
in the coding of $\dot J$ expression.}
Likewise we reduce the background mass to $M=0.974$.

\bigskip

As our simple early-stage model has less structure than
the late-stage part, we have a slightly weaker comparison to make
between the two treatments.  In Sec. \ref{Sec:P-seq} we had {\it no
freedom} in our waveform superposition comparison.  Now, for the 
earlier transition, the shape of the waveform is still fully constrained,
but the overall phase--{} or time--lag between different waveforms is 
free because our simple early-stage model doesn't include any 
information about the time and rotational offsets between the various
QC sequence slices as they are rigidly embedded in the ``astrophysical'' 
spacetime.  We will have to set this by hand from the waveforms.

\bigskip
\begin{figure}
\begin{center}
\includegraphics[width=3.2in]{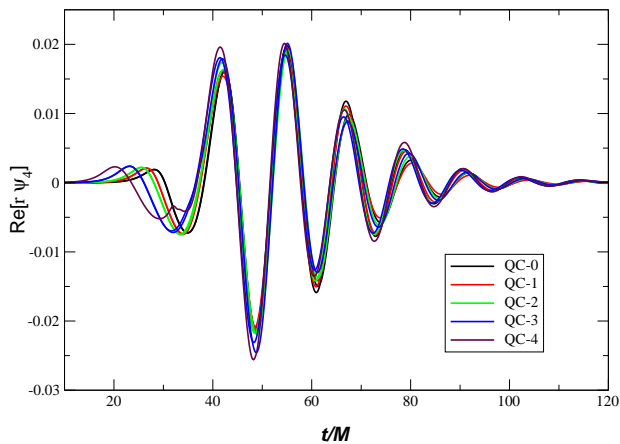}
\end{center}
\caption{Real-part of QC-sequence waveforms for the $\ell=2$ multipole.
This plot compares ten waveforms including in the FN-CL transition
time T as well as variations in the QC-sequence from QC-0 to QC-4
($L=5.0M$ to $L=7.8M$).  The agreement over a broad range of model
parameters provides support for the fidelity of our astrophysical
model.
}
\label{fig:QC-Wfs}
\end{figure}

As motivated in Sec.~\ref{Sec:P-seq}, we will filter all the waveforms 
considered here to take only the $l=2$ part.  This has the effect of 
efficiently eliminating of the higher angular frequency noise generated in the
numerical simulations, while preserving practically all the physical radiation.
Also, restricting to the $l=2,m=2$ part of the radiation as the 
overwhelmingly predominant component completely specifies the angular 
dependence so that we can, without loss of generality, focus on the 
radiation projected toward an observer on the positive z-axis.

On the basis of the waveform's phase, a good fit results from
shifting the time-axis by 0.0, -1.6, -2.9, -5.1, and -8.0 for 
the QC-0 through QC-4 waveforms, respectively.  
The results are shown in Fig.~\ref{fig:QC-Wfs}.
Table~\ref{offset} reports those values for approximate linearization
times corresponding to the FN-CL transition. One can interpret the
phase-shifts as a measure of the differential orbital-plunging times for the
elements of the $QC$ sequence.


%
\begin{table}
\caption{off set shift $t\to(t-t_0)$}
\begin{ruledtabular}
\begin{tabular}{lccc}\label{offset}
Name&	$L/M$&	$t_0/M$&$T/M$\\
\hline
QC-0&	5.0&	0.0&	9-10\\
QC-1&	5.5&	1.6&	10-11\\
QC-2&	5.9&	2.9&	10-11\\
QC-3&	6.7&	5.1&	11-12\\
QC-4&	7.8&	8.0&	13\\
\end{tabular}
\end{ruledtabular}
\end{table}

The remarkable agreement among `QC' waveforms is the result of
certainly an appropriate combination of orbital parameters $X$ and
$P$. Note that if we keep $X$ fixed and change $P$, like in the
`P-sequence', the waveforms do not superpose, and if we choose similar
$P$ and different initial separations $X$, like comparing the
waveforms of $P=2P_{ISCO}/3$ and QC-3 (see
Figs.~\ref{fig:WFP2}-\ref{fig:QC-Wfs}) or $P=5P_{ISCO}/6$ and QC-$1$
(see Figs.~\ref{fig:WFP2.5}-\ref{fig:QC-Wfs}) we do not find any
superposition of amplitudes nor phases.

It is also notable that there seems to be a pretty smooth
transition from inspiral to plunge. This suggests that it will be hard
to make a sharp definition of the ISCO, but our results compared with
those of the Post-Newtonian~\cite{Damour:2000ni,Damour:2001bu}
approximation and that of Grandclement et
al.~\cite{Grandclement:2001ed} seems to indicate that starting from
QC-3 or QC-4 data produces a good approximation to the plunge.

The striking agreement seen in Fig.~\ref{fig:QC-Wfs}
clearly supports the notion that these different initial data sets
correspond approximately to the slices of a single astrophysical spacetime.
In fact the agreement far exceeds what anyone would have expected 
from this family of conformally flat initial data sets, widely regarded
on the basis of {\em a priori} kinematical analysis
as needing to be improved upon
before any astrophysically meaningful results might be obtained.
Whatever the difficulties are with these data, they do not strongly affect 
the resulting waveforms.

\subsection{Polarization}

\begin{figure}
\includegraphics[width=3.2in]{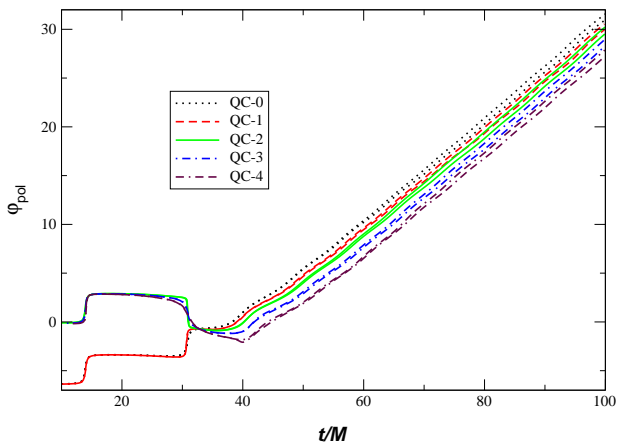}
\caption{Polarization angle.  For all QC-sequence cases the radiation 
is predominantly circularly polarized after an early period of 
initial-data dependent fluctuations.  
The colors are as labeled in Fig.~\ref{fig:QCEvsT}. 
The curves have been offset if time as
for Fig.~\ref{fig:QC-Wfs}. }
\label{fig:QCpol}
\end{figure}

\begin{figure}
\includegraphics[width=3.2in]{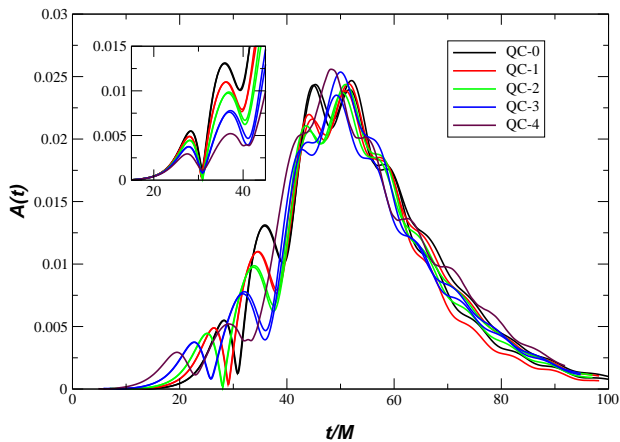}
\caption{Instantaneous radiation magnitude, independent of polarization angle.
The colors are as labeled in Fig.~\ref{fig:QCEvsT}. 
The inset shows the early part with time set to 
zero at the start of numerical evolution.}
\label{fig:QCmag}
\end{figure}

We observe an interesting correlation 
of the two polarizations of the
waveforms here represented by the real and imaginary parts of
$r\psi_4$ along the $z-$axis.
Except at early times, they are pretty much $90$ 
degrees out of phase.
The condition corresponds to circular polarization, as can be seen more  
directly if we consider our complex-valued waveforms in the representation
$\psi_4(t)=A(t) e^{i \varphi{t}}$. Physically this describes the wave according
to an instantaneous magnitude $A$ and a polarization angle $\varphi$.  
Fig.~\ref{fig:QCpol} shows the time dependence of the polarization angle
for one of our waveform cases, QC-3. After some complicated initial undulations
(after around 40M, recall that the observer is located at $r^*\approx30M$)
the polarization angle grows smoothly indicating circular 
polarization.  Fig.~\ref{fig:QCmag} shows the instantaneous magnitude of the
wave, independent of the polarization angle.  For circular polarization the 
magnitude does not vary at the radiation frequency, but more slowly.  
The figure compares wave magnitude for the QC-sequence cases.  At early times,
the wave is not circularly polarized, and the magnitude oscillates strongly 
before becoming dominated by the non-oscillating (circularly polarized) 
component.  The curves in the figure have been translated by the same 
time-shift applied 
in Fig.~\ref{fig:QC-Wfs}  
above to get phase agreement in the waveforms.
The arched circularly-polarized  late-time parts of these 
roughly superpose after this time-shift, but the
oscillatory part, shown without the time offset in the inset, is timed 
to the initial data, and varies notably as we
change among the different data sets.  This suggests that the 
non-circularly-polarized part of the waveform is predominantly 
an initial-data artifact.  Notably also, this non-astrophysical artifact
 component seems to shrink for the cases of initially more separated 
systems, as expected since the astrophysical interpretation of the 
initial data is not problematic in the limit of well-separated black holes,
reproducing two Schwarzschild black holes with an small boost.

	The physical interpretation of this polarization effect is that the
system's dynamics are dominated throughout 
the interaction by rotational motion.
Even during the ``ringing'' of the final black hole, the perturbations seem to 
circulate around the black hole rather than the bell-like ringing, one 
might naively expect.  

	We remind the reader that this discussion of waveforms is focused on
the appearance of the radiation to distant observers on the system's
rotational axis.  For observers in other locations the polarization
will generally be elliptical, but conforming to a simple pattern
consistent with the circular motion of the source.  In the equatorial
plane the observer ``sees'' no circulation in the source and the
radiation reduces to ``+'' polarization.  Moving around toward the
negative z-axis, the observed radiation approaches circular
polarization of the opposite orientation.

\subsection{Radiated Energy and Angular Momentum}

The total radiated energy $E$ and angular momentum $J$ 
at infinity per unit time 
$(u=t-r^*)$ are computed as in Ref. ~\cite{Campanelli99}, 
 \begin{eqnarray}
\frac{dE}{du}&=&\lim_{r\to\infty}\left\{ \frac{r^2}{4\pi}
\int_{\Omega}d\Omega\left| \int_{-\infty}^{u}d\tilde{u}\ 
\psi_4(\tilde{u},r,\theta,\varphi) \right|^2\right\}, \label{energy}
\end{eqnarray}
\begin{eqnarray}
\frac{dJ_z}{du}&=&-\lim_{r\to\infty}\left\{ \frac{r^2}{4\pi}\ Re\left[
\int_\Omega d\Omega\left(\partial_\varphi\int_{-\infty}^{u}d\tilde{u}
\psi_4(\tilde{u},r,\theta,\varphi) \right)\nonumber
\right.\right.
\\&& \times 
\left.\left.
\left(\int_{-\infty}^{u}du^\prime\int_{-\infty}^{u^\prime}d\tilde{u} 
\overline{\psi}_4(\tilde{u},r,\theta,\varphi)\right)\right]\right\}.  
\label{angmomentum} 
\end{eqnarray}
where $d\Omega=\sin\theta\ d\theta\ d\varphi$.

If we consider the  magnitude and polarization angle representation of 
$h$ as in Eq.\ (\ref{hcrossplus}),
{\em i.e.} $2\int\int\psi_4=A(t)e^{i \varphi(t)}$
then we find that radiation energy is generated both by time dependence in 
the magnitude $A$ and the polarization angle, but angular momentum flux is 
generated only by the rotation of the polarization vector.  Thus we can 
expect to approach maximally efficient radiation of angular momentum
$dJ=m/{\omega}dE$ \cite{Teukolsky73} for the case of circular polarization.
This is perhaps the most important consequence of the circular 
polarization pattern which we observe in our radiation.  Note that, for our 
case the most significant radiation has $m=2$ azimuthal dependence 
with a frequency near $\omega\sim0.5$ so that we can expect $dJ \sim 4 M dE$.

\subsection{Memory}
\begin{figure}
\begin{center}
\includegraphics[width=3.0in]{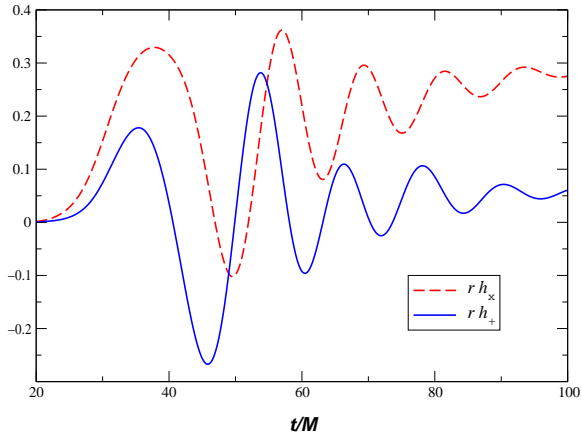}
\end{center}
\caption{$rh_+$ and $rh_\times$ for QC-3 with trivial integration constants.}
\label{fig:memory}
\end{figure}

Up to this point we have studied the radiation as represented by $\psi_4$,
which is related to the second time derivative of the usual metric 
disturbance waveform components
$h_{+}$ and $h_{\times}$ by~\cite{Teukolsky73}
\begin{eqnarray}
h_+-ih_\times&=&
2\lim_{r\to\infty}\int_0^{t}dt'\int_0^{t'}dt''\psi_4. 
\label{hcrossplus} 
\end{eqnarray}
Physically our 
waveforms correspond to the tidal acceleration of neighboring observers 
(or mirrors perhaps).  Naturally we can integrate to get their relative 
velocities and displacements, but this presents us with a problem: How do 
we set the integration constants corresponding to the initial displacements
and relative velocities of our neighboring observers?  A burst of radiation 
which allows initially quiescent observers to return to their original 
resting places is described as a ``normal'' burst, otherwise it is called a
``burst with memory.''  In trying to integrate our waveforms we find that
observers at rest when our wave arrives will not return to their original 
resting positions, and seem not to return to rest at all.  The result of 
direct integration of our waveforms is shown in Fig.~\ref{fig:memory} for a 
typical case, QC-3.

\bigskip
The figure shows a strong memory effect for  $h_{\times}$ and a much
weaker effect for $h_{+}$.  The $h_{\times}$ effect is very robust 
with respect to the the QC-sequence, while the $h_{+}$ memory effect may 
increase slightly with initial separation, but this is not clear.  
Does this mean then that colliding black holes, contrary to expectation, 
produce a burst with memory?  Not necessarily.  Our wave is only the
last part of a much longer wavetrain.  Let us adopt the hypothesis that
the full wave train is a ``normal'' burst, and consider whether our 
results contradict the hypothesis.  A key point is that neighboring
observers of the full wavetrain, which were initially quiescent, 
will not be at rest in the middle of the wavetrain, {\em where our
wave starts}.  By hypothesis, we expect these observers to return to relative 
rest so we must expect a change in their relative velocities and positions
through the passing of our part of the wave.  Our results are consistent with 
the hypothesis of finally resting observers if the ``cross'' observers, 
separated transversely by $\Delta L$ at a distance $R$ from the binary system,
are initially moving with speed 
$v\sim0.0014 \Delta L\,/R$ and the ``plus'' observers are initially 
displaced by about $0.02\Delta L\, M/R$. 
From the quadrupole formula for two orbiting point masses positioned
instantaneously on the y-axis as our black holes are initially we expect 
that the cross observers will be moving with an initial velocity of 
$2 \Omega A$ and the plus observers will be initially displaced by $A$
where $\Omega$ is the orbital velocity and $A$ is the momentary 
amplitude of the radiation.  Our results are then consistent, at an order
of magnitude level with what might be expected from extrapolated 
post-Newtonian results as presented in \cite{Blanchet:1996pi}. 
At present it would be 
inappropriate to pursue this quantitative relationship too far since there 
may be subtle effects, such as transients in the early parts of our waveforms 
or frame dragging of the radiation.
 In the end, our memory
effect is plausibly consistent with the ``normal'' burst hypothesis.  But 
this is certainly an interesting area for further study.

\bigskip
\begin{figure}
\begin{center}
\includegraphics[width=3.2in]{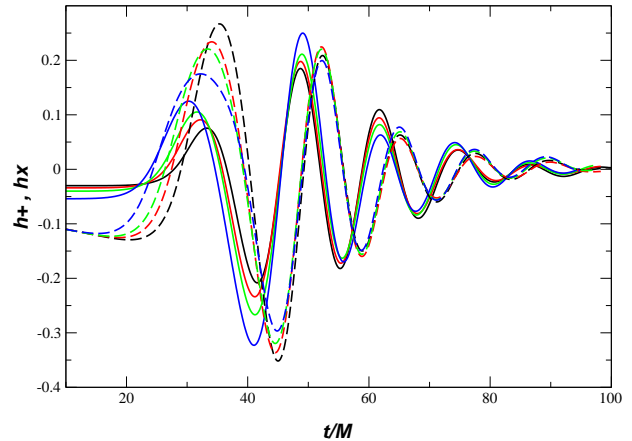}
\end{center}
\caption{$rh_+$ and $rh_\times$ with integration constants set to fit
``normal burst hypothesis''. Solid curves are '+' polarization. Dotted
curves are cross polarization.
}
\label{fig:hpluscross}
\end{figure}
If we provisionally adopt the normal burst hypothesis, it is then most
reasonable to perform the integrations required in Figure
\ref{fig:memory} with the integration constants set such that the
strains approach zero at the end of the burst.  Because our waves are
not completely quiescent and prone to small errors in the late time
region, some visual judgment is required in choosing appropriate
integration constants.  After shifting the time as above, we get
reasonable waveforms upon offsetting the $h_+$ curves by 
0.015, 0.017, 0.020 and 0.027
respectively for the cases QC-0 through QC-3 in the second integral
only.  For $h_\times$ we offset by $h_\times\rightarrow h_\times -
0.002\times t - 0.009$ 
The resulting waveforms are shown in Fig.~\ref{fig:hpluscross}.
Unlike the other plots in this section we have included only one
curve for each polarization from each QC data set.  It
is interesting to compare these ``strain'' waveforms with the other
``acceleration'' waveforms presented so far in this paper.  An obvious
difference is that the sharp growth in amplitude during the burst,
seen in Fig.~\ref{fig:QCmag} for instance, is greatly muted here.  The
stronger growth in the $\psi_4$ waveforms is a consequence of the two
factors of, $\omega$, which is rapidly increasing, obtained from two
time derivatives.  
 A consequence of this evident in the figure is that the peak in
amplitude comes earlier in the strain waveforms, pushing back to
around, $T=40M$ near the beginning of the astrophysically credible
part of the waveform.  In the early time region the figure suggests
that the large QC cases are tending toward longer wavelength early
waveform, as one would hope to see when beginning with
``astrophysically earlier'' data.  Finally we note that phase relation
between the two polarizations, which we have attributed to circular
polarization is not disrupted by two time integrations, and is equally
evident in the strain waveforms.

\subsection{Comparison with Post-Newtonian results}\label{Sec:PNcomparisons}

As a point of comparison we also consider the 3PN ISCO determined
by the method of the effective one body (E1B) of
Ref.~\cite{Damour:2001bu} and obtain the
parameters (in the ADM gauge) given in table~\ref{table:3PN}.
\begin{table}
\caption{3PN-E1B-ISCO data}
\begin{ruledtabular}
\begin{tabular}{lllllll}
Name&$L/M$&$\pm X/M$&$\pm P/M$&$J/M^2$&$M\omega$ &$m/M$\\
\hline
3PN-E1B &
$\sim7.3$ & 2.119 & 0.200 & 0.8476 & $\sim0.085$ & 0.469
\label{table:3PN}
\end{tabular}
\end{ruledtabular}
\end{table}
Note that in Ref.\ \cite{Blanchet:2001id} it was stressed the fact that
to third post-Newtonian order is not obvious that the one body
resummation is better than the `bare' results. In this case we
find that the ISCO so determined have parameter very close to
the $QC-2$ case studied here.

\begin{figure}
\begin{center}
\includegraphics[width=3.2in]{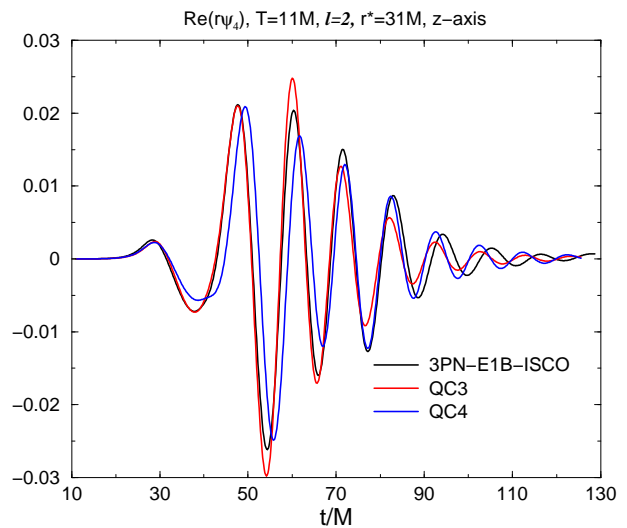}
\end{center}
\caption{Comparisons of Waveforms 3PN, QC-3, QC-4 Waveforms}
\label{fig:PNQCwfs}
\end{figure}

In order to quantify the gravitational radiation generated during
this plunge we use this 3PN ISCO parameters to evolve Bowen-York
initial data. The resulting waveform is plotted in Fig.~\ref{fig:PNQCwfs}
and compared with the QC-3 and QC-4 cases, showing closer agreement
with the QC-3 waveform.

For the full nonlinear part of the simulation
we have used a nonuniform grid of
$512^2\times256$ and $448^2\times224$ with boundaries at $37M$  We
have also considered different nonuniform grids and boundary locations
on another $512^2\times256$ simulation to check the robustness of the
results.
 The largest runs lasted
for 10 clock wall hours on the 64 nodes of the SR-8000 Hitachi computer at
the LRZ and used up to 170Gb of RAM memory.

The analysis of the PN determination of the ISCO shows
that its separation and linear momentum parameters are close to those
of the case QC-3. In retrospective, as we have seen, the plunge
waveform does not dramatically change for the five cases studies,
i.e. QC-0 to QC-4, hence our predictions for the plunge waveforms are
compatible with the third Post-Newtonian order orbital parameters
for the ISCO as well.


\section{Summary of astrophysical results}\label{Sec:Analysis}

In the last section we have presented a detailed analysis of resulting 
waveforms  from our model for the final plunge of 
a system of equal-mass, non-spinning 
(irrotational), non-elliptically orbiting binary black holes.
results.  The domain of astrophysical validity waveform is 
limited to the time from right around the first burst peak 
forward but the surprisingly lack of internal model-dependence in our
results suggests  
that we can be confident (say to the level of 20\%) in our description of
the final radiation burst from such a system.
We find these results convincing enough to merit their adoption, among
interested researchers, as a provisional description for relevant 
astrophysical work, such as in the development of strategies for
observation of gravitational waves. 
As it has been emphasized that any new information may be very valuable
for improving the efficiency of gravitational wave search algorithms
\cite{Flanagan97b}, and for the interest of readers not yet immersed in
the effort to model relativistic systems we briefly review the results 
from a more applied astrophysical viewpoint.  

Our waveforms cover the final few cycles of gravitational radiation from 
the peak onward.  Within about four cycles the signal is reduced to only a few
percent of its peak value.  In geometric units this takes about $50M$, 
which corresponds, for a $20 M_{\odot}$ system to about $0.1 ms$ or, for a
$10^6 M_{\odot}$ system of supermassive black holes about one minute.  During
this brief time vast amounts of energy are released, equivalent to about 
2.5-3\% of the system's total mass and $\sim 12\%$ of its angular momentum.
  The peak gravitational wave luminosity is about $0.0015 c^5/G$ 
or about $5\times10^{56} ergs/s$.  The radiation is strongly 
polarized in the same manner as expected from a rotating gravitational 
source with circular polarization for radiation propagating along the 
rotational axes and ``+'' polarization for radiation in the equatorial
plane.  As generally expected the radiation is predominated 
by an $l=2$, $m=2$, $s=-2$ spin-weighted spherical harmonic angular 
distribution 
so that that the intensity is 16 times greater along the axes than in 
the equatorial plane.

The simple appearance of our waveforms, as shown in Fig.~\ref{fig:QC-Wfs}
suggests that rather than asking for awkward numerical data files,
other researchers may benefit from a straightforward analytic fitting
of our result with essentially the full information at our
level of confidence.  For a representative case, the QC-3 waveform, we
can impose circular polarization, a frequency and amplitude 
dependence of the form
\begin{eqnarray}
\omega(t)&=&\Bigg\lbrace{
  \begin{array}{ll}
    \omega_0+\alpha_0 (t-t_{\omega0})&t\leq t_{\omega0}\\
    \omega_0 +\frac{(\omega_0-\omega_{QNM})(t-t_{\omega_0})}
{t_{\omega_0}-t_{\omega_1}}&t_{\omega0}<t<t_{\omega_1}\\
    \omega_{QNM}&t \geq t_{\omega_1}
  \end{array}}\\
\sigma(t)&=&\Bigg\lbrace{
  \begin{array}{ll}
    \sigma_0&t\leq t_{\sigma_0}\\
    \sigma_0+\frac{(\sigma_0-\sigma_{QNM})(t-t_{\sigma_0})}
{t_{\sigma_0}-t_{\sigma_1}}&t_{\sigma_0}<t<t_{\sigma_1}\\
    \sigma_{QNM}&t \geq t_{\sigma_1}
  \end{array}}\\
\varphi(t)&=&\varphi_0+\int_{t_0}^t{\omega(t'){\mathrm d}t'}\\
A(t)&=&e^{a_0+\int_{t_0}^t{\sigma(t'){\mathrm d}t'}}\\
r \psi_4&=&A(t)\,e^{-i \varphi(t)}. 
\end{eqnarray}
A simple interpretation of the parametrization we have chosen
for the plunge waveform starts from the amplitude-phase
representation of the waveform $r \psi_4$ as above. We then model
the time dependence of the orbital frequency of the last stage of the inspiral
by a linear increase that changes slope through the plunge
piece of the waveform and then matched the final quasi-normal frequency of the 
ringing down remnant black hole.
For the time variation of the amplitude we have chosen to match between
an initial exponential growth and a final exponential decay.

\begin{figure}
\includegraphics[width=3.2in]{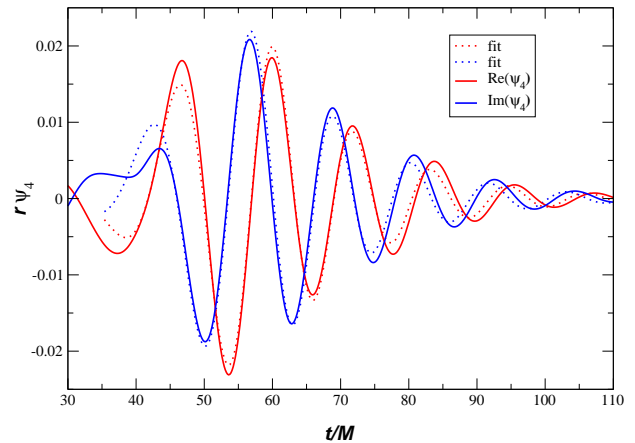}
\caption{Comparison with the fitted curve.}
\label{fig:QCfit}
\end{figure}

The quantities labeled ``QNM'' are set to their expected values for 
the least damped quasi-normal mode from black hole perturbation theory,
in this case $\omega_{QNM}=0.55$ and $\sigma_{QNM}=-0.073$.  The quantity 
$\alpha_0$ is chosen informally 
to have the value $0.0085$.  
The other eight quantities
are chosen by non-linear least-squares fit to have the approximate respective
values 
 $\{\omega0=0.2894,\ t_{\omega0}=33.00,\ t_{\omega1}=67.05,\ \sigma0=0.2192,\  
t_{\sigma0}=32.54,\ t_{\sigma1}=62.92,\ \varphi_0=-4.40,\ a_0=-6.31\}$. 
A comparison of the fit to the original curve is shown in Fig.~\ref{fig:QCfit}.
We plan to further investigate the ``fitting'' representation of the waveforms
to optimize the number and robustness of the parameters to be used 
as well as the choice of the fitting functions. This will also facilitate
the matching of the plunge waveforms with a, for instance, Post-Newtonian one
for the inspiral phase.

The curve shown for $r \psi_4(t)$ displays only the time dependence.
For the full wave information, we have
\begin{eqnarray}
\psi_4(t-r^*,\theta,\varphi)&=&{}_{-2}Y_{2\,2}\ \psi_4^{(m=+2)}(t-r^*)\nonumber\\
&&+{}_{-2}Y_{2\,-2}\ \psi_4^{(m=-2)}(t-r^*)
\end{eqnarray}
where the ${}_sY_{lm}$ are spin-weighted spherical harmonics, 
\begin{eqnarray}
{}_{-2}Y_{2\,2}&=&\sqrt{5/4\pi}\cos^4(\theta/2)\exp(2i\varphi)\\
{}_{-2}Y_{2\,-2}&=&\sqrt{5/4\pi}\sin^4(\theta/2)\exp(-2i\varphi).
\end{eqnarray}


\section{Discussion}\label{Sec:discussion}

We have produced the first astrophysical model for binary black hole
coalescence waveforms.  We use the Lazarus method applying the
techniques of numerical relativity in the strongly interacting
intermediate plunge region and close-limit black hole perturbation
theory in the late-time region.  As a stand-in for the ``far-limit''
early part of the dynamics we have taken data from a family of
solutions to Einstein's equations restricted to a spacelike slice of
spacetime. This Lazarus approach is ideal in this application for
which each component treatment is at best marginally applicable by
itself.  Not only does the Lazarus method expand the scope of
treatable problems, but it provides a natural framework for
cross-checking the various treatments against each other.

In the first part of the paper we have performed a thorough study 
of the performance the late-time treatment, including the numerical 
relativity and close-limit regions on a series of initial data set
approaching the fiducial target ``ISCO'' data, the results of which were 
recently reported on in Ref.\ \cite{Baker01b}.  the present study strongly
establishes the effectiveness of our late-time methods.

Our early-time model is based on a selection from one  
of the standard families of black hole 
initial data used in numerical relativity, the QC-sequence.
Because of the generic uncertainty in describing astrophysically accurate 
initial data consistent with Einstein's equations for strongly interacting 
black holes, we had little basis, {\em a priori}, for trusting these data 
as astrophysically reasonable.  Thus, the main part of this paper has
addressed the robustness of the overall results against variations in the
physical transition point between the early and late-time portions of our 
model.  In other words, we have tested the {\em kinematically} defined 
initial data sets against numerical relativistic {\em dynamics}.  The 
results show an overwhelming correspondence, which strongly suggests 
that, at least for, non-spinning black holes, the subtleties of selecting 
initial data are not as consequential to the resulting waveforms as 
has been generally expected in the community, and supporting our results 
as a rough first-look at the gravitational radiation which can be expected
from the coalescence of equal-mass non-spinning black holes.

We find the radiation to be dominated by circularly polarized,
$l=2,|m|=2$ waves radiating 3\% of the system's total energy in a few
cycles and significant angular momentum, so that the remnant black
hole has an angular momentum parameter of roughly $a\sim 0.7 M$, as we
have summarized the results in relation to astrophysical gravitational
wave observation observation in the last section.

In the introductory remarks, we noted the importance of having at
least a provisional astrophysical coalescence study not only for the
results it provides to observers, who need information for developing
their observational strategies, but also for the benefit of theorists
who wish to direct their effort efficiently toward the purpose of
astrophysical application. In this light, we conclude with a
characterization, suggested by the results, of the course we will
pursue in future work.  We first note that past developments in
numerical relativity have been critical in our application and have
already brought the state of the art to the point where it can be
applied astrophysically.  One of the great theoretical worries has
been that astrophysical initial data for relatively close black holes
might require a great deal of further work.  Our results suggest this
problem may not be critical, but further work is needed when the
individual spins of the interacting black holes may become relevant.
It will also be interesting to learn how generally this waveform
robustness extends when studying other initial data families.  A
weakness of our present early-time model is that it provides no direct
connection to the inspiral radiation, and thus no cross check with
post-Newtonian (PN) dynamics. Further work in this direction will
benefit from work on initial data which includes PN dynamical
radiation information, as well as longer-running numerical simulations
allowing us to extend waveforms farther back in time.  Considering the
waveforms here presented, we are impressed by its simplicity, a simple
multipolar description with monotonically increasing frequency and
smoothly changing amplitude.  However, for systems of strongly
spinning black holes or with significantly unequal masses other
arguments indicate complicated waveforms resulting from very involved
interactions.  A significant observationally motivated goal will then
be to learn how generally, in the parameter space of plausible black
hole pairs, we can expect the qualitatively simple type interaction
and radiation.  We expect such a characterization to the parameter
space to require significant additional effort, but we can begin soon
by comparatively studying co-rotational systems and exploring the
lowest order spin and unequal mass effects more quickly.

\acknowledgments 
We would like to thank Bernd Br\"ugmann for many discussions.
We also wish to thank Bernard Schutz for calling our attention on
the importance of the polarization. We are grateful to
Miguel Alcubierre and the Cactus Team lead by Ed Seidel
for help on numerical issues related to this work.  
M.C. was partially supported by a Marie-Curie Fellowship (HPMF-CT-1999-00334).
Our full numerical computations have been implemented within the
performed at AEI, NCSA, and Leibniz Rechenzentrum.

\bibliographystyle{bibtex/prsty}
\bibliography{bibtex/references}
\thebibliography{PRD}

\end{document}